\documentclass[11pt]{article}

\usepackage{amsfonts}
\usepackage{amsbsy} 
\usepackage{epsfig}
\usepackage{latexsym}
\input amssym.def
\input amssym.tex

\advance \topmargin by -\headheight
\advance \topmargin by -\headsep
\evensidemargin \oddsidemargin
\advance\hoffset by -3mm  
\advance\voffset by  8mm  
\def\bbbc{{\mathchoice {\setbox0=\hbox{$\displaystyle\rm C$}\hbox{\hbox
to0pt{\kern0.4\wd0\vrule height0.9\ht0\hss}\box0}}
{\setbox0=\hbox{$\textstyle\rm C$}\hbox{\hbox
to0pt{\kern0.4\wd0\vrule height0.9\ht0\hss}\box0}}
{\setbox0=\hbox{$\scriptstyle\rm C$}\hbox{\hbox
to0pt{\kern0.4\wd0\vrule height0.9\ht0\hss}\box0}}
{\setbox0=\hbox{$\scriptscriptstyle\rm C$}\hbox{\hbox
to0pt{\kern0.4\wd0\vrule height0.9\ht0\hss}\box0}}}}

\makeatletter
\@addtoreset{equation}{section}
\makeatother

\begin{document}
\hfuzz=100pt
\hfuzz=100pt
\title{{\Large \bf{Ricci-flat
Metrics with $U(1)$ Action and the Dirichlet
Boundary-value Problem in Riemannian Quantum Gravity and Isoperimetric
Inequalities}}}
\author{M M Akbar\footnote{E-mail:M.M.Akbar@damtp.cam.ac.uk}
\ \,\,\&
G W Gibbons\footnote{E-mail:G.W.Gibbons@damtp.cam.ac.uk}  
\\
\\
Department of Applied Mathematics and Theoretical Physics,
\\ Centre for Mathematical Sciences,
\\ University of Cambridge, 
\\ Wilberforce Road,
\\ Cambridge CB3 0WA, U.K.}
\maketitle
\begin{center}
DAMTP-2002-149 
\end{center}
\begin{abstract}
The Dirichlet boundary-value problem and isoperimetric inequalities
for positive definite regular solutions of the vacuum Einstein
equations are studied in arbitrary dimensions for the class of metrics
with boundaries admitting a $U(1)$ action.  In the case of trivial
bundles, apart from the flat space solution with periodic
identification, such solutions include the Euclideanised Schwarzschild
metrics with an arbitrary compact Einstein-manifold as the base,
whereas for non-trivial bundles the regular solutions include the
Taub-Nut metric with a $\bbbc P^n$ base and the Taub-Bolt and the
Euguchi-Hanson metrics with an arbitrary Einstein-K\"ahler base.  We
show that in the case of non-trivial bundles Taub-Bolt infillings are
double-valued whereas Taub-Nut and Eguchi-Hanson infillings are
unique. In the case of trivial bundles, there are two Schwarzschild
infillings in arbitrary dimensions. The condition of whether a
particular type of filling in is possible can be expressed as a
limitation on squashing through a functional dependence on dimension
in each case. The case of the Eguchi-Hanson metric is solved in
arbitrary dimension. The Taub-Nut and the Taub-Bolt are solved in four
dimensions and methods for arbitrary dimension are delineated. For the
case of Schwarzschild, analytic formulae for the two infilling black
hole masses in arbitrary dimension have been obtained. This should
facilitate the study of black hole dynamics/thermodynamics in higher
dimensions. We found that all infilling solutions are convex. Thus
convexity of the boundary does not guarantee uniqueness of the
infilling. Isoperimetric inequalities involving the volume of the
boundary and the volume of the infilling solutions are then
investigated. In particular, the analogues of Minkowski's celebrated
inequality in flat space are found and discussed providing insight
into the geometric nature of these Ricci-flat spaces.
\end{abstract}
\noindent
\section{Introduction}
In Riemannian quantum gravity and in the search for holographic
dualities relating the bulk gravitational physics to boundary gauge
theories one often encounters the Dirichlet problem of finding one or
more compact $(n+1)$-dimensional Riemannian manifolds
$({\cal{M}},g_{\mu\nu})$ with a given $n$-dimensional closed manifold
$(\Sigma,h_{ij})$ as the boundary such that $g_{\mu\nu}$ satisfies the
Einstein equation with a cosmological constant term or appropriate
matter fields and such that the metric $h_{ij}$ is induced on $\partial
{\cal{M}}\equiv\Sigma$. A classical solution $({\cal{M}},g_{\mu\nu})$
is then referred to as an infilling geometry for the boundary
$\Sigma$. Such solutions provide semi-classical approximations to the
path integral and are the starting point for quantum computations.

For a given boundary with a metric finding all infilling solutions,
however, is a formidably complex task even in the case of pure
gravity. However, certain simplifying features usually arise from
physical principles which make the problem more tractable. One often
assumes $h_{ij}$ to have a high degree of symmetry and invariant under
the action of some Lie group $G$ and/or to have other simplifying
features. Also, possible infilling geometries are often restricted to
the class of cohomogeneity one manifolds under the proper action of
$G$, i.e., dim(${\cal{M}}/G)=1$. The principal orbits of such a
solution will then share the topology and symmetry of
$\Sigma$. However, one can consider metric which are cohomogeneity one
under the group action of $G' \subseteq G$ provided $G'$ expands to
$G$ on $\Sigma$. For example one can fill in an $SU(2)\times
SU(2)$-invariant $S^3$ boundary with biaxial Bianchi-IX merics whose
principal orbits are in general $SU(2)\times U(1)$-invariant. With
only a cosmological constant term and in the absence of matter fields
the assumption of cohomogeneity in either case reduces the Dirichlet
problem to a set of ordinary differential equations arising from the
Einstein equation to be solved subject to the boundary data given by
the specification of $(\Sigma, h_{ij})$ and the condition of
regularity in the interior.  In many cases (for example, when $G$ or
$G'$ has a sufficient degree of symmetry) the general solution of the
set of ordinary differential equations and the manifolds over which
they can be extended, completely or partially, are known in
advance. The problem is then equivalent to the problem of the
isometric embedding of a given manifold $\Sigma$ into known manifolds
subject to the condition of regularity for the part(s) of the
manifolds enclosed by $\Sigma$.

In this paper we study the Dirichlet problem for boundaries which are
$S^{1}$-bundles over some compact manifolds. In general relativity
such boundaries often arise in gravitational thermodynamics. The
classic example is that of the trivial bundle $\Sigma\equiv S^{1}
\times S^2$. Manifolds with complete Ricci-flat metrics admitting such
boundaries are known; they are the Euclideanised Schwarzschild metric
and the flat metric with periodic identification.  It is known from
the work of York \cite{York1} that there are in general two or no
Schwarzschild solutions depending on whether the squashing (the ratio
of the radius of the $S^{1}$-fibre to that of the $S^2$-base) is below
or above a critical value. When such solutions exist, the solution of
the boundary-value problem is given by finding the $4$-geometries by
solving for the masses of the two black holes as functions of the two
radii \cite{York1}. Among other results presented in this paper, we
will show that it is possible to find analytic solutions of the
infilling Schwarzschild geometries in arbitrary dimension by using
methods not very well-known in the physics community. York's results
in 4-dimension extend readily to higher dimensions.

In the case of non-trivial bundles, the simplest example arises in
quantum cosmology in which the boundary is a compact $S^3$, i.e., a
non-trivial $S^{1}$ bundle over $S^2$. In the case of zero
cosmological constant, regular 4-metrics admitting such an $S^3$
boundary are the Taub-Nut \cite{Hawk} and Taub-Bolt \cite{Page}
metrics having zero and two-dimensional (regular) fixed point sets of
the $U(1)$ action respectively. These metrics are therefore
topologically distinct although their principal orbits share the same
topology and symmetry.  As we will see later in the paper such an
$S^3$-boundary can be filled in with a unique Taub-Nut solution and
two Taub-Bolt solutions in general. However, in either case the
boundary has to satisfy certain inequalities. Another regular metric
with non-trivial $S^{1}$-bundle boundaries (which are not
topologically $S^3$) is the Eguchi-Hanson metric in which case the
periodicity of the $S^1$-fibre is half of that in the case of an $S^3$
boundary (hence the boundary is topologically $S^{3}/{\Bbb
Z}_{2}$). This metric also has a singular orbit, i.e., an $S^2$
bolt. As we will see below such a boundary can be filled in with a
unique (or no) Eguchi-Hanson solution depending on its geometric data.

Regular cohomogeneity one Ricci-flat metrics in higher dimensions with
principal orbits that admit circle actions, i.e., metrics which provide
the generalisations to the four dimensional metrics above, are known
\cite{BB,PP}. We will discuss them and the conditions for their
regularity in detail in Section $2$ after describing the four
dimensional cases first. As we will see, possibilities proliferate as
one goes higher in dimension. Naturally to know how the 4-dimensional
picture changes in higher dimensions one seeks a method which avoids
details coming from dimensionality.  The existence and non-existence
of infilling solutions and, more importantly, the number of infilling
solutions as the boundary data is varied will be discussed in Section
3. As will be shown, despite the form of the metrics being rather
complicated functions of the radial coordinate (i.e., the coordinate
parametrizing the orbit spaces), it is possible to treat the Taub-Nut
and the Taub-Bolt metrics generically. In the case of trivial bundles
we have been able to solve for the infilling Schwarzschild geometries
in arbitrary dimension. This is described in Section 4. It is possible
to find the infilling Eguchi-Hanson metrics as well. However, the
explicit solutions for the Taub-Nut and Taub-Bolt infilling metrics
can only be found in lower dimensions using ordinary algebraic
methods. The higher dimensional solutions are discussed in Section 5
and can be solved using insights provided by the 4-dimensional
solutions.

Two classic issues in Riemannian geometry which the Dirichlet
boundary-value problem above connects us to are discussed in Section
$6$. One of them is the question of convexity (i.e., whether the
second fundamental form of the boundary which is determined by the
infilling geometry has positive eigenvalues or not) of the boundary
and its possible ramifications for quantum gravity. The other issue is
given by Minkowski's celebrated isoperimetric inequality which in
ordinary language tells us that in flat space for a given surface area
the greatest volume enclosed is that of a sphere. We find analogues of
Minkowski's inequalities for all of the above spaces and discuss them
in detail.

\section{Ricci-flat metrics admitting boundaries with $U(1)$ action}
\subsection{Four Dimensions}
In four dimensions, all Ricci-flat metrics that admit circle actions
can be obtained as special cases of the Taub-NUT metric. The Taub-NUT
metric is a two-parameter Ricci-flat metric and is invariant under the
group action of $G\equiv SU(2) \times U(1)$, i.e., biaxial Bianchi-IX
type. The Euclidean metric is usually written in the following
coordinates:
\begin{equation}
ds^2=\left(\frac{r^{2} - L^{2}}{\Delta}\right)dr^{2}+4 L^{2}
\left(\frac{\Delta}{r^{2} - L^{2}}\right) (d\psi+\cos \theta
d\phi)^{2}+(r^{2}-L^{2})(d\theta^{2}+\sin^{2}\theta d\phi^{2})
\label{TB}
\end{equation}  
where $\Delta=(r^{2}-2Mr + L^{2})$ and $L\le r <\infty$ and $M$ are
two parameters. $\theta$ and $\phi$ are the usual coordinates on
$S^{2}$ and $0 \le \psi <4 \pi/k$, $k \in {\Bbb Z}$, is the coordinate
parametrizing the $S^{1}$ fibre. For $k=1$ the period of $\psi$ is
$4\pi$ and hence the surfaces of constant $r$ are topologically
$S^{3}$.

The general form of the metric (\ref{TB}), however, is only valid for
a coordinate patch for which $\Delta \ne 0$. In general, $\Delta$ will
have two roots:
\begin{equation}
r_{\pm}=M \pm \sqrt{M^{2}-L^{2}}.
\end{equation} 
At the roots the metric degenerates to that of a round $S^{2}$, and
each such root therefore corresponds to a two-dimensional set of fixed
points of the Killing vector field $\partial/{\partial \psi}$ and
hence are singular orbits. However, for $M=L$ the roots coincide,
i.e., $r_{\pm}=L$ in which case the fixed-point set is
zero-dimensional as the two-sphere then collapses to a point. Such two- and zero-dimensional
fixed point sets have been given the names ``bolts'' and ``nuts''
respectively \cite{GH}.

For a bolt to be a regular point of the metric, the metric has to
``close'' smoothly near it, such that the subspace of $(r,\psi)$ has
the metric of ${\Bbb E}^{2}$. This can happen provided one gives
$\psi$ the appropriate period which is equivalent to imposing the
following condition \cite{Page}:
\begin{equation}
\frac{d}{dr}\left(\frac{{\Delta}}{r^{2}-L^{2}}
\right)_{(r=r_{root})}=\frac{1}{2kL}. \label{dercond}
\end{equation}
For a nut the metric (\ref{TB}) is regular for $k=1$ and approaches
the 4-dimensional flat metric near it (see below).  The coordinate $r$
ranges continuously from the nut or bolt till, in principle, another
root of $\Delta$ or to infinity. In the latter case the metric can be
defined over a complete manifold $(\tilde{{\cal{M}}},g_{\mu \nu})$
often called an instanton.
\subsubsection*{Self-dual Taub-NUT}
Setting $L=M$ in (\ref{TB}), one obtains Hawking's Taub-NUT instanton
\cite{Hawk} :
\begin{equation}
ds^2=\left(\frac{r + L}{r - L}\right) dr^{2}+4 L^{2}\left(\frac{r -
L}{r + L}\right) (d\psi+\cos \theta
d\phi)^{2}+(r^{2}-L^{2})(d\theta^{2}+\sin^{2}\theta
d\phi^{2})\label{TN}.
\end{equation}
The Riemann curvature tensor of the metric is self-dual -- $R_{\mu \nu
\rho \sigma} = \frac{1}{2} \epsilon _{\mu \nu \kappa \eta} \,R^{\kappa
\eta}_{\,\,\,\,\,\,\,\,\rho \sigma}$.
As already mentioned, the metric has a nut at $r=L$ and is regular for
$k=1$, i.e., if $\psi$ has a period of $4\pi$; the level-surfaces of
the regular metric are therefore topologically $S^{3}$ with a biaxial
Bianchi-IX metric on it -- a property that makes the metric
interesting for cosmology. Another interesting property of the metric
is that it is K\"{a}hler. In the mathematical literature, because of
its many special properties, this metric appears in many different
contexts. The metric is asymptotically flat and the complete metric,
$0 \le r <\infty$, i.e., the self-dual Taub-NUT or Taub-Nut instanton
has the topology of ${\Bbb R}^{4}$. To avoid confusion due to
divergent conventions in the literature, we will refer to this metric
(and its higher dimensional generalisations to be described in the
next section) as Taub-Nut metrics to distinguish them from the
Taub-Bolt metrics that have two-dimensional regular fixed-point sets
and will reserve the word Taub-NUT for the whole two-parameter family
which includes other regular and singular metrics.
\subsubsection*{Taub-Bolt Metric}
The Taub-Bolt \cite{Page} metric is the only other regular metric for
$k=1$, i.e., has level surfaces that are squashed spheres:
\begin{equation}
ds^2=\left(\frac{r^{2} - L^{2}}{r^{2}-2.5Lr + L^2}\right)dr^{2}+4
L^{2} \left(\frac{r^{2}-2.5Lr + L^2}{r^{2} - L^{2}}\right) (d\psi+\cos
\theta d\phi)^{2}+(r^{2}-L^{2})(d\theta^{2}+\sin^{2}\theta
d\phi^{2}).\label{Tbo}
\end{equation}
Here $r$ ranges from $2L$ to infinity. The two-dimensional fixed-point
set of the Killing vector $\partial/ \partial\psi$ is a regular bolt
as one can check from (\ref{dercond}). This is not self-dual unlike
the Taub-Nut metric although it is also asymptotically flat. Due to
the bolt, the complete metric has a different topology and is defined
over a manifold of topology $\bbbc P^{2} -\{0\}$, i.e., of $\bbbc
P^{2}$ with its nut removed.
\subsubsection*{$k=0$: the Schwarzschild Solution}
As first observed by Page \cite{Page}, for the degenerate case $k=0$,
one can obtain the Schwarzschild metric by taking the limit $k
\rightarrow 0$ and $L \rightarrow 0$ while keeping $r_{+}$ fixed
\cite{Page}:
\begin{equation}
ds^2=\left(1-\frac{2m}{r}\right)
dt^{2}+\left(1-\frac{2m}{r}\right)^{-1}dr^{2}+r^{2}(d\theta^{2}+\sin^{2}\theta
d\phi^{2}) \label{Sch}
\end{equation}
Here $t \in [0, \infty)$ and replaces the $\psi$ coordinate in the
previous two examples. The metric has a bolt singularity at $r=2m$
which can be made regular by identifying the coordinate $t$ with a
period of $8\pi m$. The radial coordinate $r$ has the range $[2 m,
\infty)$ and constant $r$ slices of the regular metric have the
trivial product topology of $S^{1} \times S^{2}$. The four-metric
therefore has the topology of ${\Bbb R}^{2} \times S^{2}$.
\subsubsection*{$k=2$: the Eguchi-Hanson Metric}
The only other regular metric in this family is the Eguchi-Hanson
metric \cite{EH}. It is obtained by defining $R^2=4(r^{2}-L^{2})$ and
then taking the limit $L\rightarrow \infty$ while keeping
$a^{2}=4(r_{+}^{2}-L^{2})$ constant \cite{Page}:
\begin{equation}
ds^2=\left(1-\frac{a^{4}}{R^{4}}\right)^{-1}dR^{2}+\frac{1}{4}
R^{2}\left(1-\frac{a^{4}}{R^{4}}\right)(d\psi+\cos \theta
d\phi)^{2}+\frac{1}{4} R^{2}(d\theta^{2}+\sin^{2}\theta d\phi^{2})
\label{Eguchi-Hanson}
\end{equation}
The metric is self-dual. However, since $\psi$ has period $2\pi$, the
level surfaces are $S^{3}/{\Bbb Z}_{2}$, i.e., ${\Bbb R}{\Bbb P}^3$
and hence the metric is asymptotically locally Euclidean and
asymptotically looks like ${\Bbb R}^4/{\Bbb Z}_{2}$. The complete
metric has the topology of $T^{*}(\bbbc P^{1})$.

\subsection{Higher Dimensional Generalisations}
In this section we briefly review the possible generalisations of the
four-dimensional metrics discussed above. All of the above metrics are
radial extensions of $U(1)$ bundles fibred over $S^{2}$ -- a compact
Einstein manifold. The bundles are non-trivial except in the case of
Schwarzschild. In higher dimensions we therefore seek Ricci-flat
metrics with similar Hopf bundle structure that would reduce to the
above metrics in four dimensions.  The case of four dimensions is
special in that all such metrics are obtained as special cases of the
Taub-NUT metric as we have seen in the previous section. However, this
fortuitous situation cannot exist in principle in higher dimensions as
cohomogeneity one metrics with principal orbits that are the
non-trivial $S^1$ bundles (i.e., the proposed generalisation of the
Taub-NUT family) and satisfying the vacuum Einstein equation can only
exist in even dimensions whereas the Schwarzschild metric can be
generalised in arbitrary dimensions (see below). The higher
dimensional generalisation of the Taub-NUT family requires the base
manifold to be K\"{a}hler (complex projective space for Taub-Nut
solutions, see below) and hence none of the metrics will be
asymptotically flat or Euclidean, locally or globally. With
Schwarzschild in higher dimensions we have more choices for the base
manifold than a sphere with the usual round metric on it due to the
proliferation of compact Einstein manifolds/metrics in dimensions
greater than two. Any compact Einstein manifold would suffice as the
base space in this case with regularity at the bolt being achieved
only through imposing correct periodicity on the
fibre-coordinate. Therefore the Schwarzschild metrics obtained as
special cases of higher dimensional Taub-NUT metrics form a subclass
of all the possible Schwarzschild metrics in that dimension. Since
$S^{2}$ is the only compact Einstein manifold in two dimensions and is
isomorphic to $\bbbc P^{1}$, in four dimensions such a coincidence is
possible.
\subsection*{Schwarzschild metric}
For the Schwarzschild metric in $(n+1)$ dimensions, one simply
replaces the trivial bundle $S^{1} \times S^{2}$ by $S^{1} \times
M_{n-1}$:
\begin{equation}
ds^{2}=\left(1-\frac{\mu}{r^{n-2}}\right)
dt^{2}+\left(1-\frac{\mu}{r^{n-2}}\right)^{-1}dr^{2}+r^{2}ds_{n-1}^{2},
\label{schd}
\end{equation}
where $\mu$ gives the black hole mass $m$ which for $M_{n-1}
\equiv S^{n-1}$ is \cite{Myers}
\begin{equation}
\mu=\frac{16\pi G\,m}{(n-1) \mathrm{Vol}(S^{n-1})}.
\end{equation}
The bolt singularity at $r^{n-2}=\mu$ can be removed by periodically
identifying the coordinate $t$ with a period
\begin{equation}
\beta_{\tau}=\frac{4\pi}{n-2}{\mu}^{\frac{1}{n-2}}.
\end{equation}
The coordinate $r$ then takes values from ${\mu}^{\frac{1}{n-3}}$ to
infinity and defines a complete metric over a manifold with ${\Bbb
R}^{2}\times M_{n-1} $ topology possessing an
$(n-2)$-dimensional fixed point set of the Killing vector $d/dt$, i.e., a
bolt. For $M_{n-1} \equiv S^{n-1}$ the metric is
asymptotically Euclidean and in general for other choices of base
manifolds are asymptotically conical with the special case $M_{n-1} \equiv S^{n-1}/\Gamma$, where $\Gamma$ is a discrete
subgroup of $SO(n+1)$, for which it is asymptotically locally
Euclidean. For a recent discussion on various possibilities of base
spaces for Schwarzschild metrics in various dimensions and their
ramifications, see \cite{Gibbons:2002pq}.
\subsection*{The Taub-NUT Family}
The generalised Ricci-flat metrics with principal orbits which are
non-trivial $S^1$ bundles were constructed in \cite{BB} and
independently in \cite{PP}. Recently they have been discussed in
\cite{Marika} and in \cite{Awad}. Such a metric is obtained by adding
a radial coordinate to the metric on the $U(1)$ bundle over a compact
homogeneous K\"{a}hler manifold $M$ of $n$ complex dimensions endowed
with an Einstein-K\"{a}hler metric ${\cal{R}}_{ij}=\lambda
{\tilde{g}}_{ij}$ and subjecting the $(2n+2)$-dimensional metric to
Ricci-flatness. The metric on the bundle is
\begin{equation}
ds^{2}_{bundle}= R^{2}\,\omega \otimes \omega+ ds^{2}_{M},
\label{bunmet}
\end{equation}
where
\begin{equation}
\omega=d\tau+A 
\end{equation}
is a connection on the bundle such that $dA$ is the the K\"{a}hler
form on $M$ with $\tau$ parametrizing the $S^{1}$ fibre. The quantity
$R^{2} > 0 $ is some function on $M$. The bundle is invariant under
the group action $U(1)\times H$ where $H$ is the symmetry of the
base. However, this is not necessarily the maximal symmetry, as we
will discuss later.

The general Ansatz for the $(2n+2)$ dimensional space is then taken by
adding a radial coordinate $r$:
\begin{equation}
ds^{2}=\gamma(r)^{2}dr^{2} +\beta(r)^{2}
\omega \otimes \omega+\alpha(r)^{2} {ds^{2}_{M}}. \label{TAB}
\end{equation}
The K\"{a}hlerian choice of $M$ renders the vacuum Einstein equations
in the simple form:
\begin{equation}
2n\beta \left(\frac{\alpha'}{\gamma}\right)' + \alpha
\left(\frac{\beta'}{\gamma}\right)'=0,
\end{equation}
\begin{equation}
2n \beta \left(\frac{\beta^{2}\gamma}{\alpha^{3}}-\frac{\beta'
\alpha'}{\beta \gamma}\right) - \alpha \left(
\frac{\beta'}{\gamma}\right)'=0,
\end{equation}
\begin{equation}
(2n-1)\left(\frac{\alpha'}{ \alpha \gamma } \right)^{2}+
\frac{1}{\gamma^{2}}
\left(\frac{\alpha'\beta'}{\alpha\beta}\right)+\left(\frac{\alpha'}{\gamma}\right)'\frac{1}{\alpha
\gamma}+2\frac{\beta^{2}}{\alpha^{4}}-\frac{\lambda}{\alpha^{2}}=0.\footnote{
Note that the third term of this equation has a typo in \cite{BB}.}
\end{equation}
Adding the first two equations and choosing the coordinate gauge in
the form
\begin{equation}
\gamma \beta = cL\ge 0, \label{gauge}
\end{equation}
one finds
\begin{equation}
\alpha^{2}=c(r^{2}-L^{2}).
\end{equation}
With this explicit form of $\alpha(r)$, $\beta(r)$ is given by the 
integral:
\begin{equation}
\beta^{2}=c\lambda rL^{2}(r^{2}-L^{2})^{-n}\left(
\int_{L}^{r}\frac{(s^{2}-L^{2})^{n}}{s^{2}}ds - C \right),
\label{beta}
\end{equation}
where $C$ is an integration constant. Note that the gauge (\ref{gauge}) is slightly
superfluous. Looking at (\ref{TAB}), it is easy to see that $c$
appears as an overall multiplicative factor. We could therefore have
absorbed $c$ into $L$. However, writing the gauge in this form would
help us recover the Taub-Nut and Taub-Bolt metrics in their familiar
forms by just setting $c$ to a constant value and setting the correct
value for $\lambda$ without having to redefine $L$, as we will see
below.

The above considerations are local and do not prescribe any
periodicity on $\tau$. As shown in \cite{PP}, to extend the above
metric globally over a manifold, as we do next, $\tau$ is required to
have a period of
\begin{equation}
\Delta \tau=\frac{4\pi\,p}{|\lambda|\, k},
\end{equation}
where $k$ is a positive integer unrestricted as yet, and $p$ is a
non-negative integer such that the first Chern class of the tangent
bundle on $H_2(M,{\Bbb Z})$ is divisible by $p$. One can now obtain
complete positive definite metrics provided one removes the possible
singularities arising from the fixed-points of the Killing vector
$\partial/\partial \tau$.
\subsubsection*{Taub-Nut}
The fixed point set of the Killing vector $\partial/\partial \tau$ would
be zero dimensional if both $\alpha$ and $\beta$ goes to zero at
$r=L$. This requires setting $C=0$ and $\tilde{r}=L$ in
(\ref{beta}). As discussed in \cite{BB,PP}, this will be a regular nut
provided the metric near the nut approaches the flat metric. This is
only possible if one choose the base manifold to be $\bbbc P^{n}$ with
the Fubini-Study metric on it. The principal orbits are then spheres
via the standard Hopf fibration of $S^{2n+1}$ and hence near $r=L$ the
metric approaches the flat metric on ${\Bbb R}^{2n+2}$, i.e.,
$d\rho^2+\rho^2\,d\Omega^{2n+1}$. Note that for any other choice of
compact K\"{a}hler base space the nut-singularity cannot be removed.
The period of $\tau$ is $4\pi(n+1)/\lambda$ as $p=(n+1)$ for $\bbbc
P^{n}$.

By setting $\lambda=2$ and $c=2$, note that the four dimensional
self-dual Taub-Nut metric Eq.(\ref{TN}) can be
reproduced.\footnote{Note that with this choice, the Fubini-Study
metric on $\bbbc P^{1}$ is
\begin{equation}
ds^{2}_{\bbbc P^{1}}=\frac{1}{2}(d \theta^{2}+\sin \theta^{2} d \phi^{2}).
\end{equation}} 
In higher dimensions the solutions are higher order polynomials in
$r$:\\
\begin{equation}
\beta_6^{2}=\frac{4}{3}\frac{L^2(r-L)(r+3L)}{(r+L)^2},
\label{beta11}
\end{equation}
\begin{equation}
\beta_8^{2}=\frac{4}{5}\frac{L^2(r-L)(r^2+4rL+5L^2)}{(r+L)^3}\label{beta2},
\end{equation}
\begin{equation}
\beta_{10}^{2}=\frac{4}{35}\frac{L^2(r-L)(5r^3+25r^2L+47rL^2+35L^3)}{(r+L)^4},
\label{beta3}
\end{equation}
\begin{equation}
\beta_{12}^{2}=\frac{4}{63}\frac{ L^2(r-L)(7r^4+42r^3L+102r^2L^2+122rL^3+63L^4)}{(r+L)^5}.
\label{beta4}
\end{equation}
One can integrate 
\begin{equation}
\beta^{2}=c\lambda rL^{2}(r^{2}-L^{2})^{-n}\left(
\int_{L}^{r}\frac{(s^{2}-L^{2})^{n}}{s^{2}}ds\right), \label{betaNUT}
\end{equation}
to obtain
\begin{equation}
\beta^{2}=\frac{c\lambda\,r\,L^2}{(r^2-L^2)^n}\left(\frac{L^{2n-1}}{\sqrt{\pi}}\Gamma(n+1)\,\Gamma(\frac{1}{2}-n)+\left(\frac{1}{r^2}\right)^{\frac{1}{2}-n}\,\frac{1}{2n-1}\,\,_2F_1[\frac{1}{2}-n,-n,\frac{3}{2}-n,\frac{L^2}{r^2}]\right).
\end{equation}
However, this expression, while exact, consists of two terms coming
from the two limits of the integral (in which the value of the
integral at the lower limit has been simplified using Gamma functions)
and hence is not very useful or illuminating for practical
purposes. However, it is possible to express $\beta^{2}$ as a single
expression which captures the simple product form of
Eqs.(\ref{beta11})- (\ref{beta4}). For this we will have to wait until
the next section where its utility will also be demonstrated.
\subsubsection*{Taub-Bolt}
For $r \ge L$ the integral
$\int_{L}^{r}\frac{(s^{2}-L^{2})^{n}}{s^{2}}ds$ is a monotonically
increasing function of $r$ starting from zero. Any positive value of
$C$ can therefore be
matched with a unique $r \equiv
r_{b}$ \cite{Marika}. Therefore, for $C >0$, we can replace Eq.(\ref{beta}) with 
\begin{equation}
\beta^{2}=c\lambda rL^{2}(r^{2}-L^{2})^{-n}\left(
\int_{r_{b}}^{r}\frac{(s^{2}-L^{2})^{n}}{s^{2}}ds\right). \label{beta1}
\end{equation}
This automatically guarantees that $r_b>L$.

The $2n$ dimensional set of fixed point of $\partial/\partial \tau$ at
$r=r_{b}$ forms a bolt -- a singular orbit. However, this will be a
regular point of the metric provided the $(\tau,r)$ sub-plane looks
like an ${\Bbb E}^{2}$ at $r=r_{b}$.  We now expand (\ref{beta1}) in
powers of $(r-r_{b})$:
\begin{equation}
\beta^{2}= \frac{c\lambda L^2}{r_{b}}\,(r-r_{b})\,\, +\,\,\, \mathrm{
higher\,\,\,order\,\,\,terms}. \label{betaint}
\end{equation}
Recalling that $\tau$ should have a period $4\pi p/|\lambda|k$, the
sub-space of $(\tau,r)$ would be flat if the bolt is located at
\begin{equation}
r_{b}=p\,L/k.
\end{equation}
Obviously this requires $k (\in {\Bbb Z})$ to be less than
$p$. Therefore there are $(p-1)$ bolt type solutions for any given
Einstein-K\"{a}hler base.

For $\bbbc P^n$ (therefore $k> (n+1)$)
\begin{equation}
r_{b}=(n+1)L/k.
\end{equation}
Note that only for $k=1$ is the asymptotic behaviour of the solution
similar to that of the Taub-Nut solution. Other $(n-1)$ bolt
solutions, which do not have any analogues in 4-dimensions, though
regular, do not have the same asymptotic behaviour.  For $k=1$, the
bolt appears at $r_{b}=2L$ in four dimensions, $r_{b}=3L$ in six
dimensions, i.e., at $r_{b}=(n+1)L$ in $(2n+2)$ dimensions. We here
mention explicit solutions for $k=1$, with subscripts indicating the
dimensions as before:
\begin{equation}
\beta_6^{2}=\frac{4}{3}\frac{(r-3L)(r+L)}{(r-L)^2},
\end{equation}
\begin{equation}
\beta_8^{2}=\frac{1}{5}\frac{(r-4L)(4r^5+16r^4 L+44r^3 L^2 +176 r^2
L^3 +764 r L^4 -5 L^5)}{(r^2-L^2)^3},
\end{equation}
\begin{equation}
\beta_{10}^{2}=\frac{4}{35}\frac{(r-5L)(5r^7+25r^6 L+97 r^5 L^2 +485
r^4 L^3 +2495 r^3 L^4 + 12475 r^2 L^5 +62235 L^6 +7
L^7)}{(r^2-L^2)^4}.
\end{equation}
In the rest of the paper, unless otherwise stated, we will use
Taub-Bolt to mean any solution for arbitrary value of $k$ (which is,
however, less than $p$). Also, it is important to note that for the
Taub-Nut solutions we require the total $(2n+2)$-dimensional space to
approach flatness at the nut which is only possible if $\bbbc P^n$ is
the base. However, in the case of bolt-type solutions, at a bolt the
metric is the product of flat ${\Bbb E}^2$ and an Einstein manifold of
constant radius and hence are regular Ricci-flat solutions for any
choice of Einstein-K\"{a}hler base. The periodicity of the
fibre-coordinate has to be adjusted in this case depending on the
value of $p$ which in turn will determine the locations of the bolt
in each of the $(p-1)$ bolt solutions. In general the periodicities
are different for different choices of base manifolds and the explicit
form of $\beta(r)^2$ will be different in each case and for each
possible value of $k$. However, Taub-Bolt would be used for all of
them in the same way Schwarzschild is used generically irrespective of
the choice of base.

Apart from the $(p-1)$ regular bolt solutions, another regular
solution exists for the degenerate case of $k=p$ by taking limits in the
same way we have obtained the Eguchi-Hanson metric in the case of four
dimensions (see below). Also at the limit $k=0$ one obtains
the Schwarzschild metric. However, these are only a subset of
Schwarzschild metrics in even dimensions. The Taub-NUT family being
even dimensional cannot offer any Schwarzschild metric in odd
dimensions. Therefore one obtains only the sub-class of Schwarzschild
metrics with Einstein-K\"{a}hler base in any even dimension from the
Taub-NUT family and any odd dimensional Schwarzschild metric is
precluded automatically.
\subsubsection*{Eguchi-Hanson}
For the degenerate case of $k=p$, the $S^{1}$-fibre has a period of
$\frac{4\pi}{\lambda}$. Metrics obtained by taking limits identical to
four dimensional case, as described in Sec. 2 (i.e., by defining the
coordinate $R^2=\lambda(r^2-L^2)$ and taking $r_{b}$ and $L$ both to
infinity while keeping $\lambda (r_{b}^2-L^2)=a^2$ a finite constant)
are regular and give the generalisations to the Eguchi-Hanson
metric. They can
most conveniently be written by choosing $\lambda=2(n+1)$. The
$(2n+2)$-dimensional metric then has the succinct form:
\begin{equation}
ds^{2}=\left(1-\frac{a^{2n+2}}{r^{2n+2}}\right)^{-1}dr^{2}+
r^{2}\left(1-\frac{a^{2n+2}}{r^{2n+2}}\right)(d\psi+A)^{2}
+r^{2}ds_{M}^{2}. \label{Egu}
\end{equation}
The bolt at $r=a$ is regular with $\psi$ having a period of
$\frac{2\pi}{n+1}$ for which the principal orbits are $S^{2n+1}/{\Bbb
Z}_{n+1}$ if one chooses $\bbbc P^n$ with the Fubini-Study metric as
the base. The metric (\ref{Egu}) is then Einstein-K\"{a}hler and was
first found by Calabi by abstract geometric methods \cite{calabi} and
was later found by directly solving the Monge-Amp\'{e}re equation in
\cite{FiedGib}. However, $M$ can be any Einstein-K\"{a}hler manifold
(see \cite{berber} and \cite{PP}).
\section{The Dirichlet Problem: Uniqueness and Non-uniqueness}
In the Dirichlet problem one seeks to obtain non-singular solutions
$({\cal{M}},g_{\mu\nu})$ which fill in a given boundary
$(\Sigma,h_{ij})$ such that ${\partial \cal{M}}= \Sigma$ and
$g_{\mu\nu}|_{\partial \cal{M}}=h_{ij}$. In our case a boundary is an
$S^{1}$ bundle over a compact Einstein manifold with $h_{ij}$ having
the form
\begin{equation}
ds^{2}_{\Sigma}=\alpha^{2} {ds^{2}_{M}} +\beta^{2} \omega \otimes
\omega, \label{bdmet}
\end{equation}
where $\alpha$ and $\beta$ -- the radii of the base manifold and the
$S^{1}$-fibre respectively -- are known quantities and constitute what
we will be referring to as the boundary data. The boundary $\Sigma$ is
invariant under the group $G\equiv U(1)\times H$ where $H$ is the
symmetry group of the base manifold $M$. However, $G$ is not
necessarily the maximal symmetry of $\Sigma$. For example when the
boundary is an $S^1$ bundle over $S^2$, $G$ may enlarge from $U(1)
\times SU(2)$ to $SU(2)\times SU(2) \, \sim SO(4)$ depending on the
values of $\alpha$ and $\beta$ and the periodicity of the
fibre-coordinate. The same is true for the higher dimensional metrics
discussed above.

In the Dirichlet problem we seek to find infilling metrics as
functions of $\alpha$ and $\beta$. Since we know the Ricci-flat
metrics that have principal orbits sharing the topology and symmetry
of $\Sigma$, our problem is equivalent to the isometric embedding
problem of $\Sigma$ into known manifolds
$({\tilde{\cal{M}}},g_{\mu\nu})$. When such an embedding is possible,
compact part(s) of $({\tilde{\cal{M}}},g_{\mu\nu})$ cut by the
hypersurface $(\Sigma,h_{ij})$ constitute a solution to the Dirichlet
problem.
\subsection{Schwarzschild Metrics}
For a boundary $\Sigma\equiv S^1 \times S^2$, the pair
$(\alpha,\beta)$ constitutes the canonical boundary data with the
interpretation that $\alpha$ represents the radius of a spherical
cavity immersed in a thermal bath with temperature $T= \frac{1}{2\pi
\beta}$. It is known from the work of York and others
\cite{Braden,York1} that for such canonical boundary data, apart from
the obvious infilling flat-space solution with proper identification,
there are in general two black hole solutions distinguished by their
masses which become degenerate at a certain value of the squashing,
i.e., the ratio of the two radii $\frac{\beta}{\alpha}$. This can be
seen in the following way.  First rewrite the Schwarzschild metric
(\ref{Sch}) in the following form:
\begin{equation}
ds^2=\left(1-\frac{2m}{r}\right)
64\pi^{2}\,m^{2}\,d\tau^{2}+\left(1-\frac{2m}{r}\right)^{-1}dr^{2}+r^{2}(d\theta^{2}+\sin^{2}\theta
d\phi^{2}) \label{Sch1},
\end{equation}
where $t=8\pi \tau$ such that $\tau$ has unit period. With this
definition one can simply read off the proper length -- alternatively
the radius -- of the $S^{1}$ fibre and that of the $S^{2}$ base. They
are:
\begin{equation}
\alpha^{2}= r^{2}
\end{equation}
and
\begin{equation}
\beta^{2}= 16 m^{2} \left(1-\frac{2m}{r}\right)
\end{equation}
It is easy to see that for a given $(\alpha, \beta)$, $r$ is uniquely
determined whereas $m$ is given by the positive solutions of the
following equation:
\begin{equation}
m^{3}-\frac{1}{2}\alpha\, m^{2}+\frac{1}{32}\alpha \beta^{2}=0. \label{eqM}
\end{equation}
By solving this equation for $m$, the two Schwarzschild infilling
geometries are found\footnote{Note that the boundary data $\beta$ is
the radius of the $S^{1}$ fibre and {\it{not}} the proper length
$\beta_{\tau}$, though they are related: $\beta_{\tau}= 2\pi
\beta$}. There are in general two positive roots of Eq.(\ref{eqM})
provided $\frac{\beta^{2}}{\alpha^{2}} \le \frac{16}{27}$. When the
equality holds the two solutions become degenerate and beyond this
value of squashing they turn complex. The remaining root of
Eq.(\ref{eqM}) is always negative. Therefore the two infilling
solutions appear and disappear in pairs as the boundary data is
varied. The explicit solutions are mentioned in \cite{York1}.

\subsubsection*{Higher Dimensions}
\begin{figure}[!ht]
  \begin{center}
    \leavevmode
    \vbox {
      \includegraphics[width=11cm,height=6.7cm]{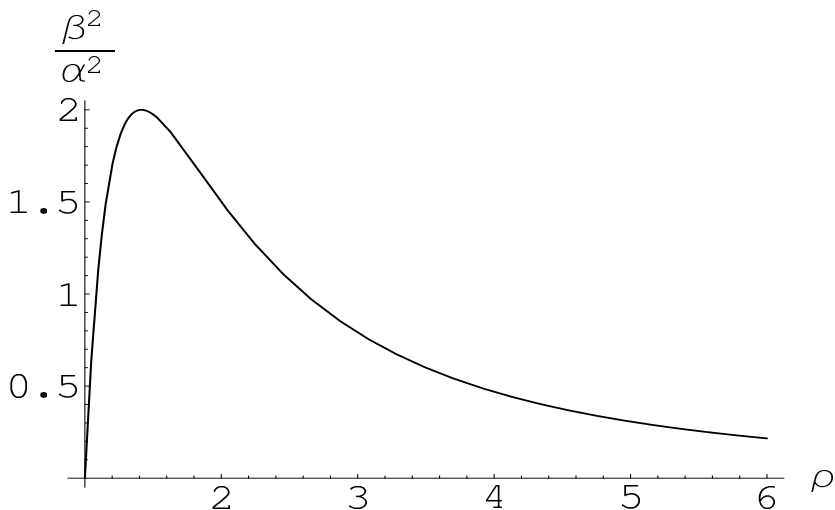}
      \vskip 3mm
      }
    \caption{Schwarzschild solutions: below a critical-value
    of the squashing ($\frac{\beta^{2}}{\alpha^{2}}$)
    there are always two distinct infilling Schwarzschild solutions
     which become
    degenerate at the critical value. The critical value and the exact
    shape of the curve depend on dimension.} 
    \label{Fig1}
  \end{center}
\end{figure}

It is not difficult to see that in arbitrary dimensions for $\Sigma
\equiv S^1\times M_{n-1}$, there will in general be two Schwarzschild
solutions or no solution. Simply note that the ratio of the two radii
as a function of $\rho \equiv r/{\mu}^{\frac{1}{n-2}} \in [1,\infty)$
is:
\begin{equation}
\frac{\beta^{2}}{\alpha^{2}}=\frac{16}{n-2}\,\rho^{-2}\left(1-\frac{1}{\rho^{n-2}}\right)\label{sqshr}.
\end{equation}
For any $n$ the general behaviour is the following:
$\frac{\beta^{2}}{\alpha^{2}}$ starts from zero and grows
monotonically to a maximum value and then decreases and approaches
zero asymptotically at infinity, as in Figure \ref{Fig1}. The maximal
value of the squashing depends on the number of dimensions only and
can be found by differentiating Eq.({\ref {sqshr}}). The condition on
the admissible boundary data in $(n+1)$ dimensions therefore reads:
\begin{equation}
\frac{\beta^{2}}{\alpha^{2}}\le \left (1/2\,n\right )^{-2\,\left
(n-2\right )^{-1}}-\left (\left (1/2 \,n\right )^{-\left (n-2\right
)^{-1}}\right )^{n}\label{sqsh}.
\end{equation}
Any squashing satisfying the inequality will occur for two values of
$\rho$ corresponding to two distinct values of $\mu$. If the squashing
of a given data $(\alpha, \beta)$ exceeds this value, there will be no
positive solution for $\mu$ and hence no real infilling Schwarzschild
solution can be found with positive mass. At the equality of
(\ref{sqsh}), the two solutions are degenerate as in four dimensions.
\subsection{Eguchi-Hanson in Arbitrary Dimension}
For the Eguchi-Hanson metric (\ref{Egu}) in $(2n+2)$ dimensions, it is
fairly straightforward to check that the infilling solutions are
unique. The squashing
\begin{equation}
\frac{\beta^2}{\alpha^2}=1-\left(\frac{1}{\rho}\right)^{2n+2}
\end{equation}
as a function of $\rho\equiv\frac{r}{a} \,\in [0,\infty)$, increases
monotonically from zero and approaches unity as $\rho \rightarrow
\infty$. For a given boundary data $(\alpha,\beta)$, the infilling
solution is given trivially by
\begin{equation}
r=\alpha
\end{equation}
and 
\begin{equation}
a=\alpha^{2n+2}\,\left ({\frac {\alpha^2-\beta^2}{{\alpha}^{2}}}\right )^\frac{1}{2n+2}.
\end{equation}
Solutions will exist for any data $(\alpha, \beta)$ provided
$\alpha>\beta$ which is apparent from the form of the metric
(\ref{Egu}).
\subsection{Taub-Nut and Taub-Bolt metrics}
\subsubsection{Four Dimensions}
For a boundary $\Sigma$ which is a twisted $S^{1}$ bundle over $S^{2}$
and endowed with the metric
\begin{equation}
ds_{3}^2=\beta^{2}(d\psi+\cos \theta
d\phi)^{2}+\alpha^{2}(d\theta^{2}+\sin\theta^{2} d\phi^{2}),
\end{equation}
where $\psi$ has a period $4\pi$, a Taub-Nut infilling would be
possible if the following system of equations admits real solutions
for $r$ and $L$:
\begin{equation}
\alpha^{2}-r^{2}+L^{2}=0, \label{eqnut1}
\end{equation}
and
\begin{equation}
\beta^{2}-4L^{2}\,\,\frac{r-L}{r+L}=0.\label{eqnut2}
\end{equation}
This has been solved in detail in \cite{mma} as a special case of
self-dual Taub-NUT-(anti)de Sitter solutions. The important
observation is that this system admits the discrete symmetry $(r, L)
\leftrightarrow (-r, -L)$, which inspires one to make the following
substitution:
\begin{equation}
\begin{array}{rcl}
x&=&r+L,\\
y&=&r-L.
\end{array}\label{e4}
\end{equation}
Eqs.(\ref{eqnut1})-(\ref{eqnut2}) then transform into the following
two bivariate equations (for $x$ and $y$):
\begin{equation}
xy-\alpha^2=0 \label{e5}
\end{equation}
and 
\begin{equation}
y x^{2}-2 x y^2 +y^3 -\beta^{2}x=0. \label{e6}
\end{equation}
The discrete symmetry, $(r, L) \leftrightarrow (-r, -L)$ is now
preserved in $(x,y) \leftrightarrow (-x,-y)$. Substituting for $x$
($y$) one can obtain a univariate equation in $y$ ($x$):
\begin{equation}
y^{4}-2\alpha^2y^2+\alpha^2(\alpha^2-\beta^2)=0.
\end{equation}
This is quadratic in $y^{2}$. The four solutions for $y$ therefore
will appear in pairs with opposite signs. Since $\alpha^{2}$ is
positive, this implies that the four corresponding solutions of
$(x,y)$ are of the form $(x_{1},y_{1})$, $(x_{2},y_{2})$ and
$(-x_{1},-y_{1})$, $(-x_{2},-y_{2})$.  Since the set of solutions have
the symmetry $(x_{i},y_i{}) \rightarrow (-x_{i},-y_{i})$, by applying
the transformation $(x,y) \rightarrow (-x,-y)$, we would obtain no new
solutions for $(x,y)$ (hence for $(r,L)$) and would reproduce the same
set. Therefore there are four points in the $\bbbc^{2}$ plane where
the two polynomials (\ref{e5})-(\ref{e6}) meet, i.e., four solutions
for $(r, L)$ which are related by the reflection symmetry $(r, L)
\leftrightarrow (-r, -L)$.

For convenience, let us rewrite $y^{2}=z$. Then the solutions are
\begin{equation}
\begin{array}{rcl}
z_{1}&=&\alpha^{2}-\alpha\beta,\\
z_{2}&=&\alpha^{2}+\alpha\beta\\
\end{array}
\end{equation}
If $\beta >\alpha$ then $z_{1}$ is negative which means that $y$ will
be imaginary and so will $x$ by virtue of equation Eq.(\ref{e5}). For
either of $z_{1}$ or $z_{2}$ to give a real solution, we must have
$r>L$. For this to happen, one can show trivially, using
Eq.(\ref{e4}), that one requires
\begin{equation}
\alpha^{2}>z_{i}.
\end{equation}
This can only be satisfied by $z_{1}$. Therefore for any boundary data
$\alpha>\beta$, one can fill in with a unique Taub-Nut metric.

For the same boundary, $(\Sigma,h_{ij})$, Taub-Bolt infilling
solutions are possible if the following system can admit real
solutions for $r$ and $L$ such that $r>2L$:
\begin{equation}
\alpha^{2}-r^{2}+L^{2}=0,
\end{equation}
and
\begin{equation}
\beta^{2}-4L^{2}\,\,\frac{r^{2}-2.5Lr+L^{2}}{r^{2}-L^{2}}=0.
\end{equation}
However, in this case the solution is not unique. This can be seen by
noting the behaviour of the squashing as a function of the variable
$\rho=r/L$:
\begin{equation}
\frac{\beta^{2}}{\alpha^{2}}=4\frac{\rho^2-2.5\rho+1}{(\rho^2-1)^{2}}
\label{bolt4}
\end{equation}
At $\rho=2$, this is zero and as $\rho$ is increased it increases to
reach a maximum and then decreases to reach zero at infinity. The
maximum value of $(\mathrm{squashing})^2$ is
$(\frac{3}{8}\,{3}^{2/3}-{\frac {9}{8}}\,\sqrt [3]{3} +1) \sim 0.1575$
and occurs approximately at $\rho \sim 2.851708133$. For any boundary
data $(\alpha,\beta)$ for which squashing is below this limit, the
$\Sigma$ can be filled in with two Taub-Bolt solutions. Note that
Eq.({\ref{bolt4}}) can be solved exactly for $\rho$ giving exact
infilling geometries as functions of the boundary data
$(\alpha,\beta)$. We do not, however, write the solutions here as they
are unwieldy and not particularly illuminating.

One must have noted a fundamental difference between the
Taub-Nut/Bolt, Eguchi-Hanson cases and the Schwarzschild case. In the
Taub-Nut/Bolt and Eguchi-Hanson cases the $S^1$ fibre of the boundary
$\Sigma$ is required to have a prescribed period in order to afford
regular infilling solutions. Any boundary with a different periodicity
of the fibre-coordinate hence cannot be filled in with {\it{regular}}
Taub-Nut/Bolt solutions irrespective of the value of boundary data. In
the case of Schwarzschild, however, the periodicity of the $S^1$-fibre
is determined by the masses of the infilling black holes. Because of
the product topology of the boundary, the periodicity of the
$S^1$-fibre of the boundary in this case is ``arbitrary'' in the sense
that one can always redefine the coordinate parametrizing the fibre as
we did in Section 3.1 and it is meaningful to talk about its
periodicity only after the solutions have been found.

\subsubsection{Higher Dimensions}
One can similarly treat the filling in problem in higher dimensions
with Taub-Nut and Taub-Bolt metrics and try to solve them
algebraically for $r$ and $L$ using the explicit forms of $\alpha(r)$
and $\beta(r)$ mentioned in Section $2.2$. It is possible to reduce
the problem to a one-variable one by treating the squashing as a
function of the variable $\frac{r}{L}$ as we have done for the
Taub-Bolt and Schwarzschild solutions (which equally could have been
adopted for the Taub-Nut in Section 3.3.1). However, the corresponding
equations soon become too difficult to tackle algebraically with
ordinary methods. We will return to the issue of explicit solutions
later in Section 5. For the present purpose, i.e., to see whether
infilling solutions are unique or non-unique and for what ranges of
the boundary data they exist, we need to take a more general approach
and avoid a case-by-case study as below. This enables us to make the
following statement for Taub-Nut and Taub-Bolt infilling geometries in
arbitrary dimensions.  \\\\ 
{\bf{Theorem:}} \emph{For a non-trivial $S^{1}$-bundle over a compact
K\"{a}hler manifold of $n$ complex dimensions with metric
(\ref{bdmet}), possible Taub-Nut and Eguchi-Hanson infillings are
unique and possible Taub-Bolt infilling, irrespective of the
periodicity of coordinate parametrizing the $S^1$-fibre, is
double-valued.}\\ \\
{\bf{Proof:}} We have already shown that the Eguchi-Hanson infilling
is unique and exists for any boundary data $(\alpha,\beta)$ provided
$\alpha>\beta$.

For the case of Taub-Nut and Taub-Bolt, perhaps the most
straightforward way is to use a combination of polynomials and
differential equations. Denote the ratio
$\frac{\beta(r)^{2}}{\alpha(r)^{2}}$ by $S(r)$:
\begin{equation}
S(r)=\frac{2rL^{2}}{(r^{2}-L^{2})^{n+1}} \int
^{r}_{r_{b}}\frac{(s^{2}-L^{2})^{n}}{s^{2}}ds \label{ratio},
\end{equation}\\
where we have set $\lambda=2$ without any loss of generality (see
Comment 2 below). Recall that $r_{b}=L$ for Taub-Nut in arbitrary
dimension and that $r_{b}=pL/k$ for regular Taub-Bolt solutions where
$k$ is an integer and less than $p$.

Rescaling $r$ by $L$, one obtains 
\begin{equation}
S(\rho)=\frac{2\rho}{(\rho^{2}-1)^{n+1}} \int
^{\rho}_{r'_{b}}\frac{(s^{2}-1)^{n}}{s^{2}}ds \label{ratio1},
\end{equation}
where $r'_{b}\equiv r_{b}/L$ and $\rho=r/L$. We now obtain a first
order differential equation for $S(\rho)$ from (\ref{ratio1}):
\begin{equation}
\rho\left(\rho^{2}-1\right)S'(\rho)+\left(\rho^{2}(2n+1)+1\right) S(\rho)-2=0,\label{flow}
\end{equation} 
where prime denotes differentiation with respect to $\rho$. Note that
this equation is inhomogeneous, first order and linear and its
solution is unique for any boundary data $\left(S(\rho),\rho
\right)$. More importantly, it is {\emph{non-autonomous}}, unlike the
Einstein equations we started with. This latter property, though not
desirable in most cases, will greatly facilitate our understanding of
the boundary-value problem under consideration.

The apparent singular point $\rho=1$ of Eq.({\ref{flow}}) is a regular
singular point as one can guess. We will discuss it further when we
deal with the nut-case. For the purpose of exposition we start with
the Bolt case.
\subsection*{Bolt case}
At a bolt, i.e, at $\rho=r'_{b}$ (where $r'_{b}>1$), $S(\rho)$ is zero trivially as $\beta=0$ ($\alpha$ is
nonzero). (The integral  
\begin{equation}
\int ^{\rho}_{r'_{b}}\frac{(s^{2}-1)^{n}}{s^{2}}ds =F[-n,-\frac{1}{2},\frac{1}{2},\rho^{2}]-F[-n,-\frac{1}{2},\frac{1}{2},{r'_{b}}^{2}]
\end{equation}
has a factor $(\rho-r'_{b})$ for arbitrary $n$, as can be checked by
expanding and factoring the (finite) hypergeometric series making the
squashing zero at the bolt.)

Since $S(\rho)$ is zero at $\rho=r'_{b}$, Eq.({\ref{flow}}) implies
immediately that $S'(\rho)>0$ and hence $S(\rho)$ will grow. It will
continue to grow monotonically till it reaches the value where
$\left(\rho^{2}(2n+1)+1\right)S(\rho)-2=0$ -- which is an
extremum. The second derivative at the extremum (or at any extremum),
\begin{equation}
S''(\rho)=-\frac{4(2n+1)}{(\rho^2-1)(2\rho^{2}n+\rho^{2}+1)}\label{secder},
\end{equation}
is negative. So $S(\rho)$ starts decreasing and we have a ``hump''.

Since the second derivative at any extremum is negative for any
$\rho>1$ irrespective of the initial data, a minimum can never occur
and hence $S(\rho)$ will decrease monotonically, i.e., $S'(\rho)$ will
always be negative after the maximum. This in turn implies that
$S(\rho)$ can never be negative, i.e., the curve cannot cross the
$\rho$-axis because that would violate Eq.({\ref{flow}}). (This
physically corresponds to the obvious fact that $\alpha(r)$ and
$\beta(r)$ are positive.)
\begin{figure}[!ht]
  \begin{center}
    \leavevmode
    \vbox {
      \includegraphics[width=11cm,height=6.7cm]{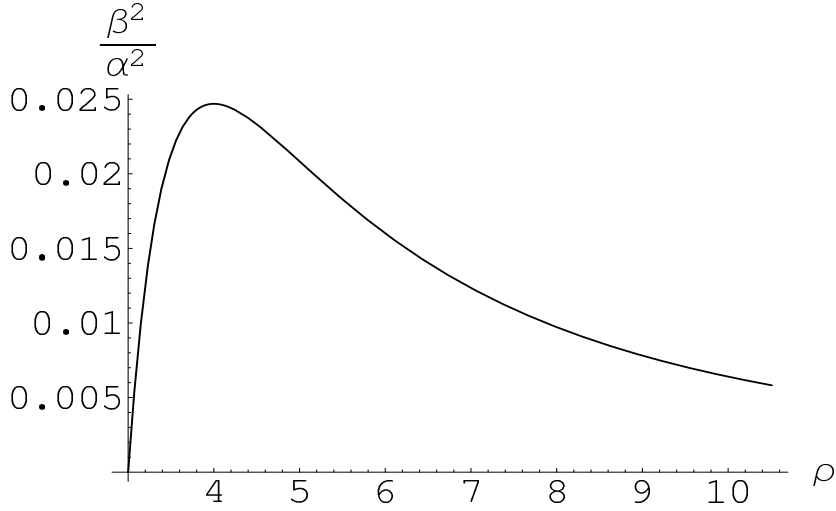}
      \vskip 3mm
      }
    \caption{Taub-Bolt solutions: below a critical-value
    of the squashing ($\frac{\beta^{2}}{\alpha^{2}}$)
    there are always two distinct infilling Taub-Bolt solutions
     which become
    degenerate at the critical value.} 
      \end{center}
\end{figure}
\begin{figure}[!h]
  \begin{center} \leavevmode \vbox {
    \includegraphics[width=11cm,height=6.7cm]{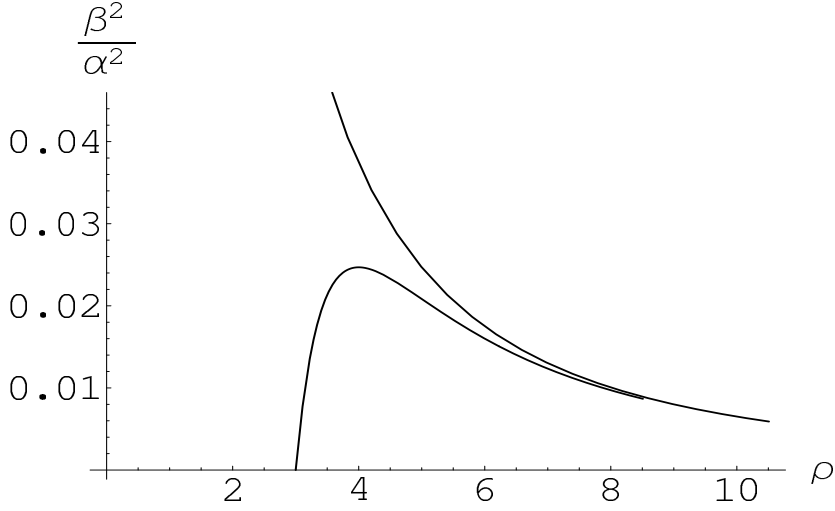} \vskip 3mm
    } \caption{Taub-Nut and Taub-Bolt solutions: any boundary $\Sigma$
    which can be filled in with Taub-Bolt solutions can necessarily be
    filled in with a Taub-Nut solution (assuming $\bbbc P^n$ and
    $k=1$). This is comparable to the case of Schwarzschild and hot
    flat space in the case $\Sigma$ is a $S^1\times S^n$.}
    \end{center}
\end{figure}

As $\rho \rightarrow \infty$, the only possibility for $S(\rho)$ is
therefore to be asymptotic to some constant value. For $\rho$ large
Eq.({\ref{flow}}) can be approximated by:
\begin{equation}
S'(\rho)+\frac{(2n+1)}{\rho}S(\rho)-\frac{2}{\rho^{3}}=0,
\end{equation}
giving 
\begin{equation}
S(\rho)=\frac{2}{(2n-1)\rho^{2}}+\frac{c}{\rho^{2n+1}}.
\end{equation}
For all $n$ and any value of $c$ (which depends on the initial data)
this approaches zero asymptotically, i.e., $S(\rho)$ is an asymptote
to the $\rho$-axis. Therefore for any boundary data $(\alpha,\beta)$
for which $\left(\frac{\beta}{\alpha}\right) <
\left(\frac{\beta}{\alpha}\right)_{max}$ there will be precisely two
infilling Taub-Bolt metrics. Physically, as $r \rightarrow \infty$,
$\beta$ stabilizes to a constant value while $\alpha$ continues to
grow linearly. Their ratio therefore approaches zero asymptotically at
infinity.
\subsection*{Nut case}
Physically $\alpha$ and $\beta$ have similar power law behaviour near
the nut and hence $S(\rho)$ approaches a constant value. This can be
seen by redefining $s=y+1$ and evaluating the integral for $y
\rightarrow 0$. One finds that $S(\rho)_{\rho=1}=\frac{1}{n+1}$
precisely. In fact, using this substitution one can show:
\begin{equation}
\int ^{\rho}_{1}\frac{(s^{2}-1)^{n}}{s^{2}}ds= \int
^{\rho-1}_{0}\frac{(y^{2}+2y)^{n}}{(y+1)^{2}}dy=
\frac{2^{n}(\rho-1)^{n+1}}{n+1}\,{\rm{Appell}}F_{1}[n+1,2,-n,n+2,1-\rho,\frac{1-\rho}{2}]
\end{equation}
where ${\rm{Appell}}F_{1}$ is the Appell hypergeometric function \cite
{bailey} of two variables, here $1-\rho$ and $\frac{1-\rho}{2}$, and
is equal to unity for $\rho=1$. Therefore, for the nut case in general
one has
\begin{equation}
S(\rho)=\frac{2^{n+1}}{(n+1)(1+\rho)^{n+1}}\,{\rm{Appell}}F_{1}[n+1,2,-n,n+2,1-\rho,\frac{1-\rho}{2}],\label{appel}
\end{equation}
which clearly has the value $\frac{1}{n+1}$ at the nut.

At the nut ($\rho=1$) one therefore has
\begin{equation}
\rho\left(\rho^{2}-1\right)S'(\rho)=0,
\end{equation}
One can verify by direct differentiation of Eq.({\ref{appel}}) that
$S'(\rho)$ has a factor $(\rho-1)$ in the denominator and the
numerator vanishes smoothly rendering the quantity
$\rho\left(\rho^{2}-1\right)S'(\rho)$ smooth at the nut and that it is
negative in general near the nut. In fact it is easier to see it in
the following way: at $\rho=1+\epsilon$, where $\epsilon$ is an
arbitrarily small positive quantity,
$\rho\left(\rho^{2}-1\right)S'(\rho)$ cannot be positive (implying
that $S'(\rho)$ cannot be positive) because this would then mean an
increase in $S(\rho)$ which is not possible since this would make the
left hand side of Eq.({\ref{flow}}) negative definite. Therefore
$S'(\rho)$ has to be negative at $\rho=1$ and $S(\rho)$ must decrease.
Employing the same argument used for the bolt case $S(\rho)$ cannot
have a minimum and will decrease monotonically to approach zero
asymptotically. Therefore for any
$\frac{\beta^2}{\alpha^2}<\frac{1}{(1+n)}$ there will be an unique
infilling Taub-Nut metric. \,\,$\Box$

Note that, for the Taub-Nut solution, $r'_{b}=1$. For the bolt
solutions, no assumption has been made about the periodicity of $\tau$
as the lower limit $r'_{b}$ of the integral (i.e., the initial value
of $\rho$) has been kept arbitrary. The only property used was that
$r'_{b}>1$ which is a necessary condition for the existence of
bolts. The infilling bolts solutions will therefore be regular
solutions if the fibre-coordinate has a period $\frac{4\pi\,
p}{\lambda\,k}$.  The periodicity of the $\tau$ coordinate and the
base manifold is determined by specifying the boundary
$(\Sigma,h_{ij})$. If the periodicity of the fibre-coordinate is such
that $r'_{b}\ne p/k$, but $r'_{b}>1$, $\Sigma$ can still be filled in
with two positive definite bolt solutions although they will be
singular at the bolt. Our analysis above applies to such possibilities
as well. Also note that we do not require the base manifold to be
$\bbbc P^n$ for the Taub-Nut case, although otherwise the solutions
will be singular at the nut.  \\ \\ {\bf{Note 1}} The flow of
Eq.(\ref{flow}) is similar for all values of $k$ though they
correspond to different geometries. Even when $r'_{b} \ne p/k$, the
pattern is unchanged as long as $r'_{b}>1$, though the corresponding
solutions are singular at their bolt. However, as one varies $r'_{b}
(>1)$ and hits $r'_{b}=1$ the flow-pattern changes abruptly. This
singular, sudden shift in the flow of Eq.(\ref{flow}) at $r'_{b}=1$
encodes the change in the {\emph{topological}} character of the
infilling solutions.  \\ \\ {\bf{Note 2}} We have found a new closed
form expression for $\beta(r)$ for the Taub-Nut through the Appell
hypergeometric functions (cf Eq.({\ref{appel}})):
\begin{equation}
\beta(r)_{Nut}^2=\frac{\lambda\,c\,\,
2^{n}\,r\,(r-L)\,L^{n}}{(n+1)(r+L)^{n}}\,{\rm{Appell}}F_{1}[n+1,2,-n,n+2,1-\frac{r}{L},\frac{L-r}{2L}].\label{appel1}
\end{equation}
Note that this form explicitly shows how the ratio of $\beta(r)$ and
$\alpha(r)$ is non-zero at the nut, though they are separately zero.
To our knowledge, the above form (\ref{appel1}) 
has not been found before. In terms of ordinary hypergeometric
functions the closed form expression for $\beta(r)_{Bolt}^2$ consists
of two terms, i.e., the difference of the function at the two limits
of the integral (\ref{betaint}) and hence is not particularly
illuminating.  \\ \\ 
{\bf{Comment 1}} Note that the boundary data for which there is an
infilling Taub-Nut solution is restricted by the squashing being less
than or equal to $\sqrt{\frac{1}{n+1}}$; beyond this there will be no
nut solution.  By setting $S'(\rho)=0$ the upper bound on the
squashing $S_{max}$ and the value of $\rho$ at which it occurs in the
case of Taub-Bolt can be found by solving Eq.(\ref{ratio1}) and
Eq.(\ref{flow}) simultaneously. Eliminating $S_{max}$, for example,
one needs to solve the following equation for $\rho$:
\begin{equation}
\left(\rho^{2}(2n+1)+1\right)\frac{2\rho}{(\rho^{2}-1)^{n+1}}
\int^{\rho}_{p/k}\frac{(s^{2}-1)^{n}}{s^{2}}ds-2=0, \label{highlim}
\end{equation} 
which, unfortunately, does not provide a simple expression for $\rho$
and consequently for $S_{max}$. However, following the previous
discussions, we know that there will be a unique positive solution to
Eq.(\ref{highlim}) which one can find for specific $n$ and $k$
numerically.  \\ \\
{\bf{Comment 2}} For a given boundary metric one can choose the
cosmological constant $\lambda$ for the base manifold arbitrarily. The
above calculations were carried out for $\lambda=2$. The statements on
the limits on squashing for the Taub-Nut and the Taub-Bolt are to be
understood in this light. Obviously the choice of $\lambda$ does not
affect the above arguments and one can always convert quantities from
one choice of $\lambda$ to another by basic algebraic
manipulations. The limits of squashing for $\lambda'$, for example,
are obtained from those for $\lambda$ by multiplying by
$\lambda'/\lambda$.
\subsection{On-shell action}
Before closing this section note that the on-shell action for any
infilling solution $({\cal{M}},g_{\mu\nu})$ is \cite{GHsur}:
\begin{equation}
I_{E}=-\frac{1}{8\pi G}\int_{\partial{\cal{M}}} d^{n}x\,\sqrt{h}K \label{on-shell}
\end{equation}
where $K$ is the trace of the second fundamental form
$K_{ij}=\frac{1}{2}\frac{\partial h_{ij}}{\partial\hat{n}}$ and $h_{ij}$ is
the metric of the boundary and $\hat{n}$ is the outward unit
normal\footnote{Note that in this convention the outward normal on
the boundary is positive.}. The bulk contribution vanishes so only the
boundary term contributes. Thus the boundary term plays a vital role
in the Euclidean approach to quantum gravity as was noted in
\cite{GHsur}. An analogous boundary term first appeared in the $3+1$
formulation in \cite{York-sur} and also in \cite{Regge1, Regge2}. For
a detailed discussion of the structure of the boundary term see
\cite{Charap}. 
\section{Black Holes in a Cavity of Arbitrary Dimension}
In this section we discuss finding solutions to the infilling
geometries for Schwarzschild in arbitrary dimension for arbitrary
boundary data $(\alpha,\beta)$. As mentioned earlier, the periodic
boundary conditions have the natural thermodynamic interpretation of a
spherical cavity immersed in a heat bath with temperature proportional
to the inverse of the radius of the fibre. As we will see below, it is
possible to find exact solutions for the two infilling Schwarzschild
geometries as analytic functions of the boundary data, or their
squashing to be more specific, in all dimensions.
 
To find the infilling geometries in $(n+1)$ dimensions, we need to
solve the analogue of Eq.(\ref{eqM}) obtained via the method used for
four dimensions:
\begin{equation}
C^{n}-\alpha^{(n-2)}C^{2}+\frac{1}{4}\alpha^{(n-2)}(n-2)^{2}\beta^{2}=0
\label{eqC},
\end{equation}
where $\mu$ has been replaced by $C^{n-2}$ for notational convenience.
As this is an equation in $C^{2}$ in odd dimensions, exact
Schwarzschild infilling solutions can be found immediately for
dimension five:
\begin{equation}
C_{\pm}=\frac{1}{\sqrt{2}}\sqrt{\alpha (\alpha \pm
\sqrt{\alpha^{2}-4\beta^{2}})}\label{mas}
\end{equation}
giving the two masses
\begin{equation}
M_{\pm}=\frac{3}{16}\,\pi\alpha (\alpha \pm \sqrt{\alpha^{2}-4\beta^{2}}).
\end{equation}
The condition on squashing is then
\begin{equation}
\frac{\beta^{2}}{\alpha^{2}} \le \frac{1}{2},
\end{equation}
and is in agreement with Eq.(\ref{sqsh}). 

Similarly, explicit solutions can be found in seven and nine
dimensions as functions of the boundary data by means of ordinary
methods. In the case of four dimensions, we already know the solutions
from the work of York \cite {York1}. In dimensions other than these,
the degree of (\ref{mas}) is five and above. Explicit solutions are
therefore not obtainable in general in terms of radicals, as we know
from Galois theory. However, as will be shown below, exact {{\it
analytic}} solutions are still possible. They involve techniques less
well-known to the physics community and hence probably were not found
before.

For convenience, we rewrite Eq.(\ref{eqC}) in the following form:
\begin{equation}
x^{n}-x^{2}+ p=0 \label{refeq},
\end{equation}
where $x \equiv C/\alpha$ and $p
=\frac{(n-2)^{2}}{4}\frac{\beta^{2}}{\alpha^{2}}$. Note that
$p$ can be positive only for $0\le x \le 1$. This is because $x$ here is just the inverse of $\rho$ in
Eq.(\ref{sqsh}). 

By the fundamental theorem of algebra, Eq.(\ref{refeq}) will always
admit $n$ complex solutions. As we have seen, two of these will be
positive as long as the squashing does not exceed the value given by
Eq.(\ref{sqsh}) which translates into the following requirement for
Eq.(\ref{refeq}):
\begin{equation}
p \le \left(\frac{2}{n}\right)^\frac{2}{n-2}\left(1 -\frac{2}{n}\right)\label{psqsh}.\label{4.6}
\end{equation}
The existence of two positive roots can be reproduced from purely
algebraic arguments. For $n$ odd, there will be at least one real root
which will be negative as the product of all roots, $(-1)^{n}\, p$, is
negative. Remembering Descartes rule of changes of sign (see, for
example, \cite{BC}), there will be either two other positive roots
(double roots counted twice) or no positive roots -- meaning all other
roots are complex. For $n$ even there can be up to two positive and
two negative roots while the rest are complex; the positivity of
$(-1)^{n}\,p$ in this case excludes the possibility of having one
positive root and one negative root. So there will be either two
positive solutions or no positive solutions for $C$ depending on the
value of $p$.
\subsection{Solutions}
The problem at hand is to find the $n$ solutions of Eq.(\ref{refeq})
and identify the two positive roots which are bound to exist for $p$
obeying (\ref{psqsh}). Things are much simpler than the general $n$-th
degree equations as Eq.(\ref{refeq}) belongs to the class of equations
\begin{equation}
x^{n}-a\,x^s+b=0 \,\,\,\, (n>s) \label{trinom},
\end{equation}
known commonly as trinomial equations. Trinomial equations have a long
history, and was studied by mathematicians starting from Lambert to
Ramanujan in both their general and restricted forms in parallel with
the general equation of degree $n$. An elaborate historical account
can be found in \cite{Bela}. A major success came with the work of
Birkeland in $1927$ which showed that the $n$-roots of the general
equation of degree $n$ can be expressed as linear combinations of
higher order hypergeometric functions of several variables
\cite{Birkeland}. As a special case to the general equation of degree
$n$, he showed that the general trinomial equation of degree $n$ with
arbitrary complex coefficients can be solved exactly in terms of
hypergeometric functions of order $n$ of {\it{one}} variable. We have
discussed the general solutions in detail in the Appendix. As far as
we know, there is no substantial reference to the general solutions of
trinomial equations in the literature available in English.

As in the Appendix, the general solutions of (\ref{trinom}) are given
 in terms of the variable
\begin{equation}
\zeta=\frac{n^{n}}{s^s(n-s)^{n-s}}\frac{b^{n-s}}{a^{n}}.
\end{equation}
The solutions fall into two separate sectors corresponding to
$|\zeta|<1$ and $|\zeta|>1$. The two sectors have different analytic
forms of solutions in terms of hypergeometric functions.  In our case,
$a=1$, $s=2$, $b=p$ and hence
\begin{equation}
\zeta=\frac{n^{n}}{4(n-2)^{n-2}}\,p^{n-2}. \label{com}
\end{equation}
Since $p=\frac{(n-2)^{2}}{4}\frac{\beta^{2}}{\alpha^{2}}$, $\zeta$ is
a quadratic function of squashing.

Recalling that we are interested only in the{\it{ positive}} roots of
Eq.(\ref{refeq}), which exist only for values of $p$ given by
(\ref{psqsh}), we need not consider $\zeta$ corresponding to values
exceeding the bound of (\ref{psqsh}). Now $\zeta$, as a function of
$p$, increases monotonically from zero (for $p=0$) to its maximal
value corresponding to the equality of (\ref{psqsh}). It is
straightforward to check that this maximal value of $\zeta$ is
identically unity. We therefore {\it{do not}} need to consider the set
of solutions corresponding to $|\zeta|>1$ to find the two positive
roots at all. Each of the positive roots of (\ref{refeq}) (and other
roots within this bound) can therefore be given by a single analytic
expression. This is not specific to Eq.(\ref{refeq}), however, as
$\zeta$ will be equal to unity when the general trinomial equation
$x^{n}-a\,x^s+b=0$ has equal roots. The condition for this reads
\begin{equation}
\frac{s^s(n-s)^{n-s}}{n^{n}}\,a^{n}=b^{n-s}.
\end{equation}
In our case, the two positive roots become equal at $p =
\left(\frac{2}{n}\right)^\frac{2}{n-2}\left(1 -\frac{2}{n}\right)$
which is precisely the condition above as one can check comparing it
with (\ref{com}).

\subsubsection{Analytic Solutions}
The $n-2$ roots $x_{i}$'s of Eq.(\ref{refeq}) are found from
Eq.(\ref{ns}) by setting $\gamma=1$ and $a=1$ and $b=p$ :
\begin{equation}
x_{i}=
(e^{{\frac{2\pi\sqrt{-1}}{n-2}}})^{i}\left(F_{0}(\zeta)+\frac{1}{n-2}
\sum_{\kappa=1}^{n-3} (e^{{\frac{2\pi\sqrt{-1}}{n-2}}})
^{-i\,n\kappa}\mu_{\kappa} (-p)^{\kappa} F_{\kappa}(\zeta)\right)\\\\
\\ (i=1,2,...,n-2) \label{sol1}
\end{equation}
in which 
\begin{equation}
\mu_{\kappa}=\frac{1}{\kappa}\frac{(\frac{1 -2
\kappa}{n-2}-1)!}{(\kappa-1)!(\frac{1 -2 \kappa}{n-2}-\kappa)!}.
\end{equation}
The remaining two roots are found from Eq.({\ref{s}):
\begin{equation}
x_{n-2+i}=\sqrt{p}\,e^{i\pi\sqrt{-1}}\left(\phi_{0}(\zeta)+\frac{1}{2}e^{i\,n\pi\sqrt{-1}}\,p^{\frac{n-2}{2}}\phi_{1}(\zeta)\right)\\\\
\\ (i=1,2).\label{sol2}
\end{equation}
The arguments of the function
\begin{equation}
F_{\kappa}(\zeta)=F \left(
\begin{array}{cccc}
a_{1,\kappa}, & \dots, & a_{n-1,\kappa}, & a_{n,\kappa}\\
b_{1,\kappa}, & \dots, & b_{n-1,\kappa}, & \zeta
\end{array}
\right)
\end{equation}
are given by
\begin{equation}
\begin{array}{rcl}
a_{i,\kappa}&=&\frac{\kappa}{n-2}+\frac{n-i}{n}-\frac{1}{n(n-2)}
\,\,\, (i=1,2,\dots,n),\\
b_{i,\kappa}&=&\frac{\kappa}{n-2}+\frac{3-i}{2}-\frac{1}{2(n-2)}
\,\,\, (i=1,2),\\
b_{i,\kappa}&=&\frac{\kappa}{n-2}+\frac{i-2}{n-2}+\frac{\delta_{i}}{n-2}
\,\,\, (i=3,\dots,n-1),
\end{array}
\end{equation}
where
\begin{equation}
\delta_{i}=0, {\mathrm{when}}\,\,\, i<n-\kappa,\,\,\delta_{i}=1,
{\mathrm{when}}\,\,\, i \ge n-\kappa\label{delta}.
\end{equation}
The arguments of the function
\begin{equation}
\phi_{\kappa}(\zeta)=F\left(
\begin{array}{cccc}
d_{1,\kappa}, & \dots, & d_{n-1,\kappa}, & d_{n,\kappa}\\
e_{1,\kappa}, & \dots, & e_{n-1,\kappa}, & \zeta
\end{array}
\right)
\end{equation}
are given by
\begin{equation}
\begin{array}{rcl}
d_{i,\kappa}&=&\frac{\kappa}{2}+\frac{i-1}{2}+\frac{1}{2\,n} \,\,\,
(i=1,2,\dots,n),\\
e_{i,\kappa}&=&\frac{\kappa}{2}+\frac{i}{n-2}+\frac{1}{2\,(n-2)}
\,\,\, (i=1,2,\dots,n-2),\\
e_{i,\kappa}&=&\frac{\kappa}{2}+1+\frac{i-n}{2}+\frac{\delta_{i}}{2}
\,\,\, (i=n-1),
\end{array}
\end{equation}
and $\delta_{i}$ is given through $(\ref{delta})$.
\subsubsection*{\it The two masses}
It is not difficult to single out the two positive roots from
(\ref{sol1}) and (\ref{sol2}). They are simply the $(n-2)$-th and
$n$-th roots. The masses of the two black hole solutions are therefore
given by the following simple expressions:
\begin{equation}
x_{+}=F_{0}(\zeta)+\frac{1}{n-2} \sum_{\kappa=1}^{n-3} \mu_{\kappa}
(-p)^{\kappa} F_{\kappa}(\zeta)\label{m1}
\end{equation}
and
\begin{equation}
x_{-}=\sqrt{p}\,\left(\phi_{0}(\zeta)+\frac{1}{2}
p^{\frac{n-2}{2}}\phi_{1}(\zeta)\right)\label{m2}.
\end{equation}
The expressions (\ref{m1}-\ref{m2}) converge for $\zeta<1$. For the $\zeta=1$ case,
corresponding to $p=\left(\frac{2}{n}\right)^\frac{2}{n-2}\left(1
-\frac{2}{n}\right)$, the double-root is much simpler:
\begin{equation}
x=\left(\frac{2}{n}\right)^{\frac{1}{n-2}}.\label{lesmas}
\end{equation}
The solutions (\ref{m1}) and (\ref{m2}) gives us the masses of the two
infilling black hole solutions as analytic functions of the boundary
data. This enables us to find the corresponding Euclidean actions and
other thermodynamics quantities as analytic functions of the boundary
data and sets the ground for any future study in higher dimensional
black holes in a thermal cavity. Note that, the smaller mass
(\ref{m1}) will be less than the value of $x$ given by (\ref{lesmas}})
for which the specific heat capacity
\begin{equation}
C_{A}=4\pi(n-2) \,C^{n-1} \,(1-x^{n-2})\left(\frac{n}{2} x^{n-2}-1\right)^{-1}
\end{equation}
is negative and hence the black hole solution is thermodynamically
unstable. The corresponding negative mode was found numerically in
\cite{GR}. This solution is therefore an instanton. However, the
larger mass black hole solution is locally thermodynamically
stable. 
\subsubsection{Action and Free energy}
Note that Eq.(\ref{4.6}) gives the critical temperature $T_{c}$ -- above
which the two black hole solutions exist -- to be inversely proportional to
the cavity-radius:
\begin{equation}
T_c=\frac{1}{4\pi}\left(\frac{n}{2}\right)^{\frac{1}{n-2}}\sqrt{n(n-2)}
\,\frac{1}{\alpha}.
\end{equation}
Recall that any cavity can be filled in with a unique hot flat
space 
\begin{equation}
ds^2=d\tau^2+dr^2+r^2d\Omega_{n-1}^2
\end{equation}
for any temperature $T$ by giving $\tau$ the period $\frac{1}{T}$. For
$T<T_{c}$, the only classical solution within the cavity is hot flat
space whereas for temperature equal or above $T_{c}$ the two black
hole solutions add to the list. For either of the two black hole
solutions the Euclidean action (\ref{on-shell}) in the `background' of
the periodically identified flat space is given by:
\begin{equation}
I_{E}=\frac{1}{8G}\,{\frac {{\alpha}^{n-1}\,x\left
(n\,{x}^{n-2}+2\,(n-1)\left(\sqrt {1-{x}^{n-2}}-1\right)\right)}{\left
(n-2\right )}} .\label{ACTION}
\end{equation}
The flat space action has been subtracted so as to make the action
zero for zero mass. The actions for the two black holes can be found
by direct substitution of the two positive roots of Eq.(\ref{refeq})
in Eq.(\ref{ACTION}). The action is zero for $x=0$ (by virtue of the
above subtraction) and it increases monotonically to reach a maximum
and then decreases monotonically to the minimum value
\begin{equation}
I_{E_{min}}=
\frac{\alpha^{n-1}(1-3n)}{8G(n-2)}
\end{equation}
corresponding to $x=1$. This is strictly negative and hence the action
passes through zero for some non-zero value of $x$. It is fairly
straightforward to work out that this zero occurs for
\begin{equation}
x = \left ({\frac
{4(n-1)}{{n}^{2}}}\right )^\frac{1}{\left (n-2\right )}. \label{subs}
\end{equation}
Therefore for 
\begin{equation}
1\ge x >\left ({\frac
{4(n-1)}{{n}^{2}}}\right )^\frac{1}{\left (n-2\right )},
\end{equation}
the action will always be negative. Note that the values of $x$ within
this range is always greater than the value of $x$ given by
(\ref{lesmas}) and hence the action will be negative only for the
larger mass black hole and for a restricted range, i.e., if the
temperature is sufficiently high enough. Since the free energy is
$F=\beta I_{E}$, the larger-mass black hole nucleated spontaneously
from hot flat space for any temperature above this will be globally
thermodynamically stable. For a given cavity of radius $\alpha$, this
temperature can be found using Eq.(\ref{refeq}):
\begin{equation}
T_{s}=\frac{1}{4\pi}\frac{n-2}{\sqrt{\left ({\frac
{4(n-1)}{{n}^{2}}}\right )^\frac{2}{\left (n-2\right )}-\left ({\frac
{4(n-1)}{{n}^{2}}}\right )^\frac{n}{\left (n-2\right )}}}\,\frac{1}{\alpha}.
\end{equation}
\subsubsection{Series Expansions}
The expressions (\ref{m1}) and (\ref{m2}) are exact and can be used
for precise calculation using the known properties of the higher order
hypergeometric functions (see, for example, \cite{bailey,
Slater}). However, in many situations, approximations via series
expansions in powers of the parameter are useful along with the exact
solutions. Obtaining such series for (\ref{m1}) and (\ref{m2}) is
nontrivial through their direct expansions. However, such series can
be obtained directly from (\ref{trinom}) by use of Lagrange
expansion. If
\begin{equation}
y=a+h\, \phi (y),
\end{equation}
where $\phi(y)$ a is function of $y$, the expansion of any function
$f(y)$ of $y$ is given by
\begin{equation}
f(y)=f(a)+ h(\phi f') + \frac{k^2}{2!} (\phi^2 f')'+\frac{k^3}{3!}
(\phi^3 f')''+ \dots. \label{exp}.
\end{equation}
There are three fundamental power series for all roots of a general
trinomial equations \cite{Eagle}, each of them precisely corresponding
to the three analytic expressions described in the Appendix, two for
$\zeta <1$, and one for $\zeta >1$. Since our solutions lie within the
bound $\zeta \le1$, only two of them will be of relevance.  We will
not discuss the general expansion for the general trinomial equation
(\ref{trinom}) -- interested readers are referred to
\cite{Eagle}. However, the method discussed below is the same for any
trinomial equation.

Note that for very small $p$, Eq.(\ref{refeq}) has two roots
approximately lying on a circle of radius $\sqrt{p}$ and the other
$(n-2)$ roots approximately on a circle of radius $1$ in the
$\bbbc$-plane. Following \cite{Eagle}, we rewrite Eq.(\ref{refeq}) as
\begin{equation}
y=1-p\,y^{\frac{-2}{n-2}},
\end{equation}
wherein $y\equiv x^{n-2}$. The Lagrange expansion (\ref{exp}) of the
function $f(y)=y^{\frac{1}{n-2}}$ then gives us the required series
for (\ref{m1}):
\begin{equation}
x_{+}=1-{\frac {p^{\frac{1}{2}}}{1!\,(n-2)}}-{\frac {\left (n+1\right
){p}^{2}}{2!\, \left (n-2\right )^{2}}}-{\frac {\left (n+3\right
)\left (2\,n+1 \right ){p}^{3}}{3!\,\left (n-2\right )^{3}}}-{\frac
{\left (n+5 \right )\left (2\,n+3\right )\left (3\,n+1\right
){p}^{4}}{4!\,\left (n-2 \right )^{4}}}-\dots . \label{ser1}
\end{equation}
Similarly defining $y\equiv x^2$ and rewriting Eq.(\ref{refeq}) as
\begin{equation}
y=p+y^{\frac{n}{2}},
\end{equation}
the Lagrange expansion of $f(y)=y^{\frac{1}{2}}$ gives us the
series for (\ref{m2}):
\begin{equation}
x_{-}=p^{\frac{1}{2}}+\frac{ p^{\frac{n-1}{2}}
}{1!\,2}\,+\frac{ \left (2n-1\right )p^{\frac{2n-3}{2}}  }{2!\,2^2}\,+\frac{  \left (3n-1
\right )\left (3n-3\right )p^{\frac{3n-5}{2}}  }{3!\,2^3}\,+
\frac{ \left (4n-1\right )\left (4n-3\right )\left (4n-5
\right )p^{\frac{4n-7}{2}}  }{4!\,2^4}\,+\dots  \label{ser2}
\end{equation}
The other complex/negative roots are obtained by multiplying
(\ref{ser1}) by the $(n-1)$ roots of $z^n-1=0$ and multiplying
(\ref{ser2}) by the other root of $z^2-1=0$, i.e., by $-1$
respectively.

Note that as one takes the boundary to infinity keeping the
temperature finite, the stable solution disappears and the instanton
solution $x_{-}$ survives thus making hot space unstable to the
nucleation of a black hole with negative specific heat as was seen in
\cite{Gross}. This happens for any non-zero temperature. Because this
solution has a negative specific heat it will be in unstable
equilibrium with its thermal environment. Only within a finite cavity
is it therefore possible to have a black hole which is
thermodynamically stable. However, the situation is different in the
presence of a negative cosmological constant as was shown in
\cite{BCM} for four dimensions. Work on higher dimensions are under
progress and will be reported in a forthcoming publication.
\section{Solving Taub-Nut and Taub-Bolt in Arbitrary Dimension}
In this section, we briefly discuss possible analytic solutions for
Taub-Nut and Taub-Bolt infilling geometries in higher dimensions. (We
have already mentioned the explicit solutions in the case of
Eguchi-Hanson metric in arbitrary dimension in the course of proving
uniqueness.) We have found solutions for the Taub-Nut and Taub-Bolt
solutions in four dimensions although in the case of the latter we did
not write down the explicit form as it is not particularly
illuminating. Note that higher dimensional Taub-Nut metrics all have
the same symmetry $(r,L) \leftrightarrow (-r,-L)$. With the same
substitutions used in four dimensions, namely,
\begin{equation}
\begin{array}{rcl}
x&=&r+L,\\
y&=&r-L,
\end{array}
\end{equation}
one can reduce the problem to the study of a univariate equation -- a
cubic in six dimensions, quartic in eight dimensions, i.e., an
equation of degree $(n+1)$ in $(2n+2)$ dimensions. The boundary-value
problem for Taub-Nut therefore can be solved exactly in up to eight
dimensions using radicals. We do not, however, write them here for
lack of space. For dimensions ten and above the relevant equations
will be quintic and above, and, as already mentioned, solutions in
terms of radicals are not possible in general. As discussed in the
previous section, one can solve the general equation of degree $n$ in
terms of the higher order hypergeometric functions\footnote{Recently
this has been done using $\cal{A}$-hypergeometric functions
\cite{Stu}.}. However, unlike the case of Schwarzschild, where we had
a trinomial equation, these will be higher order hypergeometric
functions of several variables.  Although it is possible in principle
to work them out explicitly, the solutions would not be illuminating
as in the case of Schwarzschild solutions and therefore we do not
attempt to do it here. Rather it is much easier to treat them
numerically. A numerical treatment is rather straightforward as we can
treat everything as a function of squashing only -- a simplification
resulting from the vanishing of the cosmological constant term.
\section{Convexity of the Solutions and Isoperimetric Inequalities}
Often the condition of convexity is applied to eliminate degenerate
infilling solutions for a given boundary. For example, an $S^{3}$ of
3-radius $r$ embedded in an $S^{4}$ of 4-radius greater than $r$
divides the $S^{4}$ into two unequal hemispheres both of which are
infilling solutions {\emph{a priori}}.  However, the smaller
hemisphere is the one which is convex, i.e., the eigenvalues of the
second fundamental form $K_{ij}=\frac{1}{2} \frac{\partial
h_{ij}}{\partial n}$ are positive (using the convention that the
outward normal $n$ is positive) whereas for the larger hemisphere it
is the opposite. Discarding the latter leaves us with one unique
infilling solution.

It is fairly straightforward to check explicitly that $K_{ij}$ has
positive eigenvalues for both of the Schwarzschild solutions and for
the unique Eguchi-Hanson solution.  For the Taub-Nut or Taub-Bolt, the
second fundamental form can easily be computed in the orthonormal
frame:
\begin{equation}
\begin{array}{rcl}
K_{\hat{1}\hat{1}}&=&\frac{1}{2\gamma}\frac{dh_{\hat{1}\hat{1}}}{dr}=\frac{\beta
r}{L},\\
K_{\hat{i}\hat{i}}&=&\frac{1}{2\gamma}\frac{dh_{\hat{i}\hat{i}}}{dr}=\frac{\beta}{4L}(\beta^{2})',\\
\end{array}
\end{equation}
where $\hat{i}>1$. Since the scale factors, $\alpha(r)$ and $\beta(r)$
both increase monotonically (for any value of $L$) with $r$, both
$K_{\hat{1}\hat{1}}$ and $K_{\hat{i}\hat{i}}$ are strictly
positive. Therefore the unique infilling Taub-Nut solution and the two
Taub-Bolt infilling solutions are all convex without imposing any
further restrictions on the boundary data than those needed for the
solutions to exist.
\subsection{Lower Bound to the Action}
Since all the solutions are convex, it follows immediately from a
theorem by Reilly \cite{Reilly} that the following inequality holds
\begin{equation}
\frac{A^{2}}{ V} > \frac{2n+2}{2n+1}\int_{\Sigma} K dA \label{reilly}
\end{equation}
where $A$ and $dA$ are the volume and the volume element of the
boundary $\Sigma$ and $V$ is the volume of the $(2n+2)$-dimensional
manifold with the boundary $\Sigma$. Since the boundaries above are
convex for any infillings the right hand side of (\ref{reilly}) is
positive for all infilling solutions above. Note that this term is
proportional to the Euclidean action (\ref{ACTION}) with a negative
proportionality constant and hence the action is negative for all
infillings. The inequality (\ref{reilly}) therefore provides a lower
bound for the action of any infilling solution.
\subsection{Minkowski's Inequality}
For a $(d-1)$-dimensional closed surface $\Sigma$ in ${\Bbb E}^{d}$,
one has the following inequality due to Minkowski (see, for example,
\cite{hilbert}):
\begin{equation}
\frac{A^d}{V^{d-1}} \ge d^{d-1}\,\,\mathrm{Vol(S^{d-1})} \label{bound}
\end{equation}
where $A$ is the $(d-1)$-volume of $\Sigma$, often referred to as
``area'', and $V$ is the volume of the ${\Bbb E}^{d}$ enclosed by it.
This states that for a closed surface of constant area the greatest
volume enclosed is that of a sphere (for which the equality holds). In
three dimensions, this gives the celebrated formula
\begin{equation}
\frac{A^3}{V^{2}} \ge 36\pi.
\end{equation}
Naturally, one therefore seeks the Ricci-flat counterparts to this
flat-space inequality. This is what we do in the following section
with our Ricci-flat metrics which admit a $U(1)$ action. It is not
obvious whether such inequalities would obey the bound (\ref{bound})
in general or under special conditions. As we will see, the machinery
we developed in the preceding sections for finding uniqueness or
non-uniqueness of the infilling solutions has already set the ground
for such an investigation. As before, we deal with four dimensions
first before going to higher dimensions. For the sake of convenience
we adopt the terminology of the flat space by denoting the
codimension-one volume of $\Sigma$ as $A$ and often refer to it as the
``area''. The term volume and $V$ will be reserved for the infilling
solutions.
\subsubsection{Schwarzschild Solutions}
\subsubsection*{Four Dimension}
For the Euclidean Schwarzschild metric Eq.(\ref{Sch}) the area of the
3-surface $\Sigma$ at radius $r$ is
\begin{equation}
A=r^2\,\int_{0}^{8\pi M}\left(1-\frac{2M}{r}\right)^{\frac{1}{2}} dt
\int_{0}^{\pi }\sin\theta \,d\theta \int_{0}^{2\pi}d\phi=32 \pi^2 M
r^2\,\left(1-\frac{2M}{r}\right)^{\frac{1}{2}}.
\end{equation}
The volume $V$ of the Euclidean Schwarzschild metric bounded by
$\Sigma$ is:
\begin{equation}
V=\int_{0}^{8\pi M} dt \int_{2M}^{r}s^2\, ds \int_{0}^{\pi }\sin\theta
\,d\theta \int_{0}^{2\pi}d\phi=\frac{32}{3}\pi^2 M
r^3\,\left(1-\left(\frac{2M}{r}\right)^3 \right)
\end{equation}
The dimensionless ratio is therefore
\begin{equation}
S\equiv \frac{A^4}{V^3}=32 \times
27\,\pi^2\,\frac{M}{r}\frac{\left(1-\frac{2M}{r}\right)^{2}}{\left(1-(\frac{2M}{r})^3
\right)^3}.
\end{equation}
Rewriting $\frac{2M}{r}\equiv x$, as we did in Section $4$, we obtain
\begin{equation}
S(x)=16 \times 27\,\pi^2\,x\frac{\left(1-x\right)^{2}}{\left(1-x^3
\right)^3} \label{S1}
\end{equation}
Recall that for a given hypersurface $\Sigma$, there are in general
two Euclidean Schwarzschild infilling metrics which are given by the
two positive solutions for $x$ satisfying the following algebraic
equation
\begin{equation}
x^3-x^2+p=0,
\end{equation}
where $p=\frac{1}{4}\frac{\beta^2}{\alpha^2}$. This gives two black
hole solutions for a given $(\alpha, \beta)$ which determines the area
$A$ of $\Sigma$. As we have shown, for the smaller-mass black hole,
which is an instanton, $x_{-}\in[0,\frac{2}{3}]$ and for the
larger-mass black hole which is thermodynamically/dynamically stable
$x_{+}\in[\frac{2}{3},1]$.

It is easy to check that $S(x)$ is a monotonically increasing function
of $x$ in the interval $[0,1)$ and blows up at $x=1$. Therefore the
isoperimetric inequality for the lower-mass unstable Schwarzschild
infilling solution is
\begin{equation}
S(x_{-})\le S\left(\frac{2}{3}\right), 
\end{equation}
i.e.,
\begin{equation}
\frac{A^4}{V^3} \le 32 \times
\left(\frac{27}{19}\right)^3\,\pi^2.
\end{equation}
For the larger mass, stable infilling solution this is
\begin{equation}
S(x_{+}) \ge S\left(\frac{2}{3}\right),
\end{equation}
i.e.,
\begin{equation}
\frac{A^4}{V^3} \ge 32 \times
\left(\frac{27}{19}\right)^3\,\pi^2.
\end{equation}
\subsubsection*{Arbitrary Dimension}
For convenience rewrite the Euclidean Schwarzschild metric in $(n+1)$
dimensions as
\begin{equation}
ds^{2}=\left(1-\left(\frac{C}{r}\right)^{n-2}\right)
dt^{2}+\left(1-\left(\frac{C}{r}\right)^{n-2}\right)^{-1}dr^{2}+r^{2}ds_{M}^{2} \label{schd1}.
\end{equation}
This metric is regular provided $t$ has a period of
$\frac{4\pi}{n-2}C$. The area of a constant $r$-slice $\Sigma$ is
given by
\begin{equation}
A=\mathrm{Vol(M)} \,\frac{4\pi \,C}{(n-2)}\,
r^{n-1}\,\left(1-\left(\frac{C}{r}\right)^{n-2}\right)^{\frac{1}{2}}
\end{equation}
and the volume enclosed by it is 
\begin{equation}
V=\int_{0}^{\frac{4\pi \,C}{n-2}} dt \int_{C}^{r}s^{n-1}\, ds  \,\,\,\mathrm{Vol(M)}=\mathrm{Vol(M)}\,\frac{4\pi \,C}{n(n-2)}\left(r^n-C^n\right)
\end{equation}
giving the ratio $S(x)$ which is therefore
\begin{equation}
S(x) \equiv \frac{A^{n+1}}{V^n}=\frac{4\pi
\,n^n}{(n-2)}\,\,\mathrm{Vol(M)}\,\frac{x(1-x^{n-2})^{\frac{n+1}{2}}}{(1-x^n)^n},
\label{Schrat}
\end{equation}
where $x\equiv \frac{C}{r}$ as before. Recall that in $(n+1)$
dimensions $x$ is the two positive roots of the trinomial equation
\begin{equation}
x^n-x^2+p=0,\label{618}
\end{equation}
where $p=\frac{(n-2)^2}{4}\frac{\beta^2}{\alpha^2}$. We found that for the
smaller-mass solution
$x_{-}\in
[0,\left(\frac{2}{n}\right)^{\frac{1}{n-2}}]$ and for the larger-mass stable
solution $x_{+} \in [(\frac{2}{n})^{\frac{1}{n-2}},1]$. 

It can easily be seen from (\ref{Schrat}) that $S(x)$ increases
monotonically for $x$ within $[0,1)$ and blows up at $1$. Therefore
the isoperimetric inequalities for the two infilling Schwarzschild
black holes in $(n+1)$ dimensions are:
\begin{equation}
\frac{A^{n+1}}{V^n} \le
\frac{4\,\pi}{(n-2)}\,\,\mathrm{Vol(M)}\,\frac{2^{\frac{1}{n-2}}(\frac{1}{n})^{\frac{n}{2}}(n-2)^{\frac{n+1}{2}}}{\left(1-(\frac{2}{n})^{\frac{n}{n-2}}
\right)^n}. \label{Schineqn1}
\end{equation}
for the smaller-mass unstable solution and 
\begin{equation}
\frac{A^{n+1}}{V^n} \ge
\frac{4\,\pi}{(n-2)}\,\,\mathrm{Vol(M)}\,\frac{2^{\frac{1}{n-2}}(\frac{1}{n})^{\frac{n}{2}}(n-2)^{\frac{n+1}{2}}}{\left(1-(\frac{2}{n})^{\frac{n}{n-2}}
\right)^n}\label{Schineqn2}
\end{equation}
for the larger mass stable solution.  Note that in the convention used
here the metric on the $(n-1)$ dimensional base manifold satisfies the
Einstein equation with a cosmological constant term of $(n-2)$ and
hence for $\mathrm{Vol(M)} \equiv S^{n-1}$, with the canonical round
metric on it, the volume is that of the ``unit'' $(n-1)$-dimensional
sphere and is equal to $2\pi^{\frac{n}{2}}/\Gamma(\frac{n}{2})$. With
this choice of base manifold it is easy to see that Minkowski's
inequality (\ref{bound}) is always true for the stable larger-mass
Schwarzschild solution in any dimension. However, it does not hold in
general for the lower-mass, unstable (instanton) solution.

\subsubsection{Taub-Nut}
\subsubsection*{Four dimension}
For the Taub-Nut metric (\ref{TN}) the area $A$ of a constant
$r$-slice is
\begin{equation}
A=32\pi^2\,L\left(r-L\right)^\frac{3}{2}\left(r+L\right)^\frac{1}{2}
\end{equation}
and the volume enclosed within it is
\begin{equation}
V=2L\,\int_{L}^{r}(s^2-L^2)\,ds \int\, d\psi \wedge \sin\theta\,
d\theta \wedge d\phi= \frac{32}{3}\,\pi^2\,L\,(r-L)^2\,(r+2L).
\end{equation}
Therefore
\begin{equation}
S(r)\equiv \frac{A^4}{V^3}=32 \times
27\,\pi^2\,L\,\frac{(r+L)^2}{(r+2L)^3}.
\end{equation}
Rewriting $\frac{r}{L}\equiv \rho$ we obtain
\begin{equation}
S(\rho)=32 \times 27\,\pi^2\,\frac{(\rho+1)^2}{(\rho+2)^3}.
\end{equation}
This is a monotonically decreasing function of $\rho \ge 1$. Therefore
for the Taub-Nut space the isoperimetric inequality is
\begin{equation}
\frac{A^4}{V^3} \le 128\,\pi^2. \label{TNequ}
\end{equation}
Obviously this does not obey Minkowski's inequality. In fact this is
exactly the {\it{opposite}} of Minkowski's inequality.
\subsubsection*{Arbitrary dimension}
The higher dimensional Taub-Nut metric has the form
\begin{equation}
ds^{2}=\gamma(r)^{2}dr^{2} +\beta(r)^{2}
\left(d\tau+A\right)^2+\alpha(r)^{2} {ds^{2}_{M}}. \label{TAB1}
\end{equation}
In $(2n+2)$ dimensions the area of the hypersurface $\Sigma$ at $r$ is
\begin{equation}
A=\mathrm{Vol(M)}\,\beta\,\alpha^{2n}\,\int \,d\tau
\end{equation}
Following the conventions adopted in Section $2.2$, here $\gamma
\,\beta=c\,L$ and $\alpha^2=c(r^2-L^2)$. Recall that $\tau$ has period
$\beta_{\tau}=4\pi(n+1)/\lambda$ (as $k=1$ for Taub-Nut), where
$\lambda$ is the cosmological constant of the base $\bbbc P^{n}$. This
gives
\begin{equation}
A=\frac{4\pi(n+1)}{\lambda}\mathrm{Vol(M)}\,(\sqrt{c}L)^{2n+1}\,\tilde{\beta}\,(\rho^2-1)^n
\end{equation}
in which $\tilde{\beta}\equiv \beta/\sqrt{c}L$ is a function of
$\rho\equiv r/L$.  The volume bounded by $\Sigma$ is
\begin{equation}
V\equiv \int \,d\tau \,\int_L^{r}\gamma \,\beta \,\alpha^{2n} \, dr \,
\mathrm{Vol(M)}=(\sqrt{c}L)^{2n+2}\,\times
\frac{4\pi(n+1)}{\lambda}\mathrm{Vol(M)}\,\int_1^{\rho} (\rho^2-1)^n\,
d\rho .
\end{equation}
The ratio is therefore
\begin{equation}
S(\rho)\equiv\frac{A^{2n+2}}{V^{2n+1}}=\frac{4\pi(n+1)}{\lambda}\mathrm{Vol(M)}\,\tilde{\beta^2}^{n+1}(\rho^2-1)^n
\left(\frac{(\rho^2-1)^n}{\int_1^{\rho} (s^2-1)^n ds}\right)^{2n+1}.
\end{equation}
It is easy to evaluate the integral in the denominator above
\begin{equation}
\int_1^{\rho} (s^2-1)^n ds=\frac{2^n}{n+1}\,(\rho-1)^{n+1}\,
_2F_{1}[1+n,-n,n+2,\frac{1-\rho}{2}]. \label{nice}
\end{equation}
This gives
\begin{equation}
\frac{A^{2n+2}}{V^{2n+1}}=\frac{4\pi(n+1)}{\lambda}\left(\frac{n+1}{2^n}
 \right)^{2n+1}\mathrm{Vol(M)}\, \,\tilde{\beta^2}^{n+1}(\rho^2-1)^n\,
 \left(\frac{(\rho+1)^n}{(\rho-1)\,_2F_{1}[1+n,-n,n+2,\frac{1-\rho}{2}]
 } \right)^{2n+1}.
\end{equation}
Recalling
\begin{equation}
\tilde{\beta}^2=\frac{\lambda\,2^{n}\,\rho\,(\rho-1)}{(n+1)(1+\rho)^{n}}\,{\rm{Appell}}F_{1}[n+1,2,-n,n+2,1-\rho,\frac{1-\rho}{2}].\label{app1}
\end{equation}
we obtain after simplification
\begin{equation}
\begin{array}{rcl}
\frac{A^{2n+2}}{V^{2n+1}}&=&4\pi(n+1)\,\lambda^n\left(\frac{n+1}{2^n}
 \right)^{n}\mathrm{Vol(M)}\,\rho^{n+1} (\rho+1)^{n(n+1)}\\ \\
 &\times&\frac{\left({\rm{Appell}}F_{1}[n+1,2,-n,n+2,1-\rho,\frac{1-\rho}{2}]\right)^{n+1}}{\left(_2F_{1}[1+n,-n,n+2,\frac{1-\rho}{2}]\right)^{2n+1}}.
\end{array}
\end{equation}
It is easy to check that the fraction involving the two hypergeometric
functions is equal to unity at $\rho=1$ and monotonically decreases
with $\rho$.  It falls faster than $\rho^{n+1}(\rho+1)^{n(n+1)}$ and
falls more sharply with increasing (integer) values of $n$.  The
isoperimetric inequality is therefore
\begin{equation}
\frac{A^{2n+2}}{V^{2n+1}} \le 4\pi(n+1)\lambda^n\,\left(\frac{n+1}{2^n}
 \right)^{n}\mathrm{Vol(M)} \,
2^{n(n+1)}\,\,
\end{equation}
The volume of $\bbbc P^{n}$ with the Fubini-Study metric satisfying
the Einstein equations with cosmological constant $2(n+1)$ is $
\mathrm{Vol(S^{2n+1})/2\pi}(\equiv \pi^n/n!)$ and hence
\begin{equation}
\frac{A^{2n+2}}{V^{2n+1}}\le 2^{2n+1}\,(n+1)^{2n+1}\mathrm{Vol(S^{2n+1})}\label{TNequ1}
\end{equation}
which is exactly the opposite of Minkowski's inequality (\ref{bound})
in $(2n+2)$ dimensions:
\begin{equation}
\frac{A^{2n+2}}{V^{2n+1}}\ge 2^{2n+1}\,(n+1)^{2n+1}\mathrm{Vol(S^{2n+1})}.\label{Minkh}
\end{equation}
Therefore for the Taub-Nut metric in arbitrary dimension the analogue
of the Minkowski's inequality is exactly the opposite of the
Minkowski's inequality in flat space. Since the Taub-Nut metrics
approach flatness near the nut, the equality of (\ref{TNequ1})
coincides with that of (\ref{Minkh}) in flat space.
\subsubsection{Taub-Bolt}
\subsubsection*{Four dimension}
For the Taub-Bolt metric in four dimensions (\ref{Tbo}) $\psi$ has a
period of $4\pi$ and hence
\begin{equation}
A=32\pi^2\,L\left(r^{2}-2.5Lr +
L^2\right)^{\frac{1}{2}}(r^2-L^2)^{\frac{1}{2}}
\end{equation}
and
\begin{equation}
V=2L\,\int_{L}^{r}(s^2-L^2)\,ds \int\,\sin\theta\, d\psi \wedge
d\theta \wedge d\phi= \frac{32}{3}\,\pi^2\,L\,(r-L)^2\,(r+2L).
\end{equation}
giving
\begin{equation}
S(\rho)\equiv \frac{A^4}{V^3}=32 \times
27\,\pi^2\,\frac{\left(\rho^{2}-2.5\rho +
1\right)^2(\rho^2-1)^2}{(\rho-1)^6\,(\rho+2)^3}.
\end{equation}
where $\rho=r/L$ as before. At $\rho=2$, $S(\rho)$ is zero and
increases monotonically to a maximum and then decreases monotonically
to zero. The maximum value is approximately $64.69449106\pi^2$ and
occurs at $\rho=5.279392752$.

Recall that the two infilling Taub-Bolt solutions are separated at
$\rho \sim 2.851708133$, i.e, for a given boundary data below a
certain squashing there will be a solutions with $\frac{r}{L} \in
(2,2.851708133)$ and another solution with $\frac{r}{L}\in
(2.851708133, \infty)$. For a given boundary, therefore, the
smaller-$\rho$ solution (corresponding to larger $L$) will satisfy the
following inequality
\begin{equation}
\frac{A^4}{V^3} \le S(2.851708133) \sim 38.29964761\, \pi^2.
\end{equation}
However, for the larger-$\rho$ (corresponding to smaller value of $L$)
solution the inequality is
\begin{equation}
\frac{A^4}{V^3} < 64.69449106\,\pi^2\,\,\,\, (\mathrm{approximately}).
\end{equation}
Minkowski's inequality,
\begin{equation}
\frac{A^4}{V^{3}} \ge 128 \pi^2 \label{Mink4d},
\end{equation}
is not satisfied by either of the solutions under any circumstances.
\subsubsection*{Arbitrary dimension}
For the $(2n+2)$-dimensional Taub-Bolt solution
\begin{equation}
ds^{2}=\gamma(r)^{2}dr^{2} +\beta(r)^{2}
\left(d\tau+A\right)^2+\alpha(r)^{2} {ds^{2}_{M}} \label{TAB2}
\end{equation}
the area of a hypersurface $\Sigma$ at constant $r$ is
\begin{equation}
A=\mathrm{Vol(M)}\,\beta\,\alpha^{2n}\,\int \,d\tau=\frac{4\pi
p}{\lambda\,k}\mathrm{Vol(M)}\,(\sqrt{c}L)^{2n+1}\,\tilde{\beta}\,(\rho^2-1)^n
\end{equation}
in which $\tilde{\beta}\equiv \beta/\sqrt{c}L$ is a function of
$\rho=r/L$. Since $\beta_{\tau}=4\pi p/k\lambda$ ($k<p$), the volume
enclosed within $\Sigma$ is
\begin{equation}
V= \int \,d\tau \,\int_{pL/k}^{r}\gamma \,\beta \,\alpha^{2n} \, dr \,
\mathrm{Vol(M)}=(\sqrt{c}L)^{2n+2}\,\times \frac{4\pi
p}{\lambda\,k}\mathrm{Vol(M)}\, \int_{pL/k}^{\rho} (\rho^2-1)^n\,
d\rho . \label{arbbolt}
\end{equation}
Here, as in the case of Taub-Nut, we are using the conventions of
Section 2.2.: $\gamma \,\beta=c\,L$ and $\alpha^2=c(r^2-L^2)$.  The
ratio is therefore
\begin{equation}
\frac{A^{2n+2}}{V^{2n+1}}=\frac{4\pi
p}{\lambda\,k}\mathrm{Vol(M)}\,\tilde{\beta^2}^{n+1}(\rho^2-1)^n
\left(\frac{(\rho^2-1)^n}{\int_{p/k}^{\rho} (s^2-1)^n
ds}\right)^{2n+1} \label{bb}
\end{equation}
The integral in the denominator above can be evaluated. However,
unlike Eq.(\ref{nice}) the resulting form is not very
illuminating. Following the example of the 4-dimensional Taub-Bolt
above, we know that a closed form expression does not exist. This is
because the equality of the isoperimetric inequality in this case does
not lie at the bolt. However, note that the behaviour does not change
in higher dimensions. 
At the bolt, $\tilde{\beta}$ is zero while others are non-zero and
therefore $\frac{A^{2n+2}}{V^{2n+1}}$ is zero unlike in the nut
case. 
As $\rho$ is increased $\frac{A^{2n+2}}{V^{2n+1}}$ will increase to a
certain value and then will start decreasing and approach zero
monotonically. The value of $\rho$ at which the hump of
$\frac{A^{2n+2}}{V^{2n+1}}$ occurs is greater than
the value of
$\rho$ for which the hump of $\frac{\beta(\rho)}{\alpha(\rho)}$ occurs. This has been shown explicitly above
for four dimensions leading to two different inequalities for the two
solutions. Note that compared to the Minkowski's inequality these two inequalities are opposite in
type. This happens also for Schwarzschild which also
possesses bolts -- however, in that case explicit inequalities are obtainable. Finally, note that for different Taub-Bolt solutions
(corresponding to values of $k\ne 1$) the corresponding
inequalities are {\it{not}} the ones obtained by dividing the
inequalities corresponding to $k=1$ by $k$. This is because the locations of
the bolts are different for different $k$'s. They need to be treated
separately should one look for the isoperimetric inequalities of a
certain type albeit they can only be found numerically.
\subsubsection{Eguchi-Hanson}
\subsubsection*{Four dimension}
For the Eguchi-Hanson metric (\ref{Eguchi-Hanson}) the 3-volume of a
constant $R$-slice hypersurface is
\begin{equation}
A=\pi^2\,R^3\left(1-\frac{a^{4}}{R^{4}}\right)^{\frac{1}{2}}
\end{equation}
and the 4-volume enclosed within it 
\begin{equation}
V=\frac{1}{8} \int_{a}^{R} s^3\,ds\, d\psi \wedge \sin\theta\, d\theta
\wedge d\phi= \pi^2\int_{a}^{R} s^3\,ds=\frac{\pi^2}{4}\left(R^4-a^4\right),
\end{equation}
giving
\begin{equation}
\frac{A^4}{V^3}=64\,\pi^2 \, \frac{1}{\left(1-\frac{a^4}{R^4}\right)}.
\end{equation}
The isoperimetric inequality is then
\begin{equation}
\frac{A^4}{V^3}\ge 64\,\pi^2.
\end{equation}
This sits precisely in the middle of the extremes of flat space and
Taub-Nut. Note that all three spaces are self-dual. It would be
interesting to investigate the isoperimetric inequalities in other
self-dual spaces in four dimensions which, however, is beyond the scope of this
paper.
\subsubsection*{Arbitrary dimension}
Recall that the Eguchi-Hanson metric (\ref{Egu}) was written using the
convention that $\lambda=2(n+1)$ and hence $\tau$ has a period of
$\frac{2\pi}{(n+1)}$.  The area of the constant-$r$ hypersurface and
the volume enclosed by it are respectively
\begin{equation}
A=r^{2n+1}\sqrt{1-\frac{a^{2n+2}}{r^{2n+2}}}\,\frac{2\pi}{(n+1)}
{\mathrm{Vol{(M)}}}
\end{equation}
and
\begin{equation}
V=\frac{1}{2(n+1)}\,
r^{2n+2}\left(1-\frac{a^{2n+2}}{r^{2n+2}}\right)\,\frac{2\pi}{(n+1)}
{\mathrm{Vol{(M)}}}.
\end{equation}
and hence
\begin{equation}
\frac{A^{2n+2}}{V^{2n+1}}=\frac{2\pi}{(n+1)}
{\mathrm{Vol{(M)}}}\,2^{2n+1}\,(n+1)^{2n+1}\frac{1}{\left(1-\frac{a^{2n+2}}{r^{2n+2}}\right)^n}
\end{equation}
giving the isoperimetric inequality for the Eguchi-Hanson metric to be
\begin{equation}
\frac{A^{2n+2}}{V^{2n+1}} \ge 2^{2n+1}\,(n+1)^{2n}
{\mathrm{Vol{(M)}}}\label{Eguhidim}.
\end{equation}
For $M\equiv \bbbc P^{n}$ which in this convention has the volume of
the unit sphere $S^{2n+1}$ divided by $2\pi$, the isoperimetric
inequality ({\ref{Eguhidim}}) reads
\begin{equation}
\frac{A^{2n+2}}{V^{2n+1}} \ge 2^{2n+1}\,(n+1)^{2n}
{\mathrm{Vol({S^{2n+1}})}}\label{Eguhidim1}.
\end{equation}
This is just $\frac{1}{n+1}$ times the $(2n+2)$ dimensional
Minkowski's inequality (\ref{Minkh}) in flat space. This is therefore
always somewhere in the middle of flat space and Taub-Nut. Note that
this inequality illustrates another fundamental difference between the
Eguchi-Hanson case and the Schwarzschild and Taub-Bolt cases than
found in terms of number of infilling solutions. In the latter two
cases $S(\rho)$ is zero at the bolt whereas in this case $S(\rho)$
blows up at the bolt. The lower bound of (\ref{Eguhidim1}) comes from
$\rho\rightarrow \infty$. This explains (\ref{Eguhidim1}) immediately
via the periodicity of the fibre.
\section{Conclusion}
In this paper we studied the Dirichlet problem for cohomogeneity one
Ricci-flat metrics whose principal orbits are $S^1$ bundles over
compact Einstein spaces. We then investigated the subsequent
isoperimetric inequalities. In the case of trivial bundles the
Ricci-flat solutions are the Schwarzschild metrics for arbitrary
choices of the compact Einstein base. In the case of non-trivial
bundles the base spaces are required to be Einstein-K\"ahler and
solutions exist only in even dimensions. The resulting Ricci-flat
solutions can be classified using their 4-dimensional analogues: the
Eguchi-Hanson, the Taub-Nut and the Taub-Bolts. All of these metrics
can be topologically classified according to the presence and absence
of singular orbits, i.e., bolts or nuts. With the correct periodicity
of the $S^1$-fibre, bolts and nuts can be made regular and hence can
be included into the complete metric. In the case of Taub-Nut one
further requires the Einstein-K\"{a}hler base to be complex projective
space.

When the boundary $\Sigma$ is a non-trivial bundle of dimension three,
it is possible to find explicit 4-dimensional Taub-Nut and Taub-Bolt
solutions exactly by treating the problem as a system of two bivariate
equations. However, these equations become rather complicated as one
goes higher in the ladder of dimensionality making a case-by-case
study rather difficult and impossible analytically. This problem can
be circumvented by using a general approach involving differential
equations and polynomials which together provide a unified way of
treating the Taub-Nut and all Taub-Bolts (including those for which
$p<k<1$) in arbitrary dimension. This method also makes their
topologically different characters rather distinct. Such an approach
is possible because Ricci-flatness effectively reduce the problem to
the one-variable problem of squashing, i.e., involving the ratio of
the two radii of $\Sigma$ rather than their separate absolute values.
The Taub-Nut infilling solution is unique and the Taub-Bolt infilling
solution is double-valued in all dimensions and do not depend on
details like the choice of base and the periodicity of the $S^1$
fibre. The upper limit on the squashing for the boundary $\Sigma$ to
admit a Taub-Nut infilling is given by a simple function of
dimension. In the case of Taub-Bolts, however, we are not so
lucky. Upper limits on squashing, although a function of dimension
only need be worked out for each dimension (and for each possible
values of $k$) separately which is a straightforward task however. In
the case that $\Sigma$ is a trivial bundle the infilling Schwarzschild
geometry is double-valued in arbitrary dimension and does not depend
on the choice of the base manifold. The upper limit on the squashing
for $\Sigma$ to admit Schwarzschild solutions is a simple function of
dimension. One may therefore guess that the origin of non-uniqueness
is geometric and is related to the presence of a singular orbit of the
group action, i.e., a bolt. This is indeed the case. However, the
presence of a bolt acts as a necessary condition only. It is not a
sufficient condition as we found that the possible Eguchi-Hanson
infilling geometry in arbitrary dimension is unique despite the
presence of a bolt. It is worth recalling that the Eguchi-Hanson
metric in four dimensions has a self-dual Riemann tensor. Hence the
set of ordinary differential equations arising from the Einstein
equations is reduced to a first order set and the various scale
factors (here the two radii of the evolving hypersurface) fix the
values of their derivatives uniquely. They, however, are required to
satisfy the constraint equation. The Einstein equation then guarantees
a unique evolution (whether it is regular or non-regular at the origin
is an {\it a posteriori} issue). This is also true for the Taub-Nut in
four dimensions which is self-dual as well. In the presence of a
cosmological constant the Taub-Nut in four dimensions becomes
Taub-Nut-(anti)de Sitter which obviously does not have a self-dual
Riemann tensor. However, it has a self-dual Weyl tensor which reduces
the system to first order and thus the above comments apply. The
condition on the boundary data to have a self-dual Taub-Nut-AdS
solution has been discussed in detail in \cite{mma}. The infilling
geometries were found as exact analytic functions of the boundary data
despite the presence of a cosmological constant which involves the two
radii rather than their squashing. In the case of Taub-Bolt-AdS the
number of solutions can be as high as ten which has been shown in
\cite{Akbar1}.

In the case of trivial bundles, infilling Schwarzschild solutions were
obtained by finding the two masses of the black holes as analytic
functions of the two radii (effectively their squashing). To our
knowledge this is the first study of this kind of higher dimensional
Schwarzschild. With the relatively recent work on the negative mode of
higher dimensional Schwarzschild in a finite cavity \cite{GR}, this
provides a straightforward generalisation of the results obtained in
four dimensions. Since we have obtained the two masses of the black
holes as exact, analytic functions of the cavity radius and
temperature, our study provides a basis for further exact
semi-classical computations of black hole thermodynamics and dynamics
as all classical quantities can be evaluated from the geometry exactly
as functions of the cavity radius and its temperature. As one would
expect, the introduction of a cosmological constant would also make
the problem rather non-trivial. It is indeed the case. However, it has
been shown in \cite{BCM} that in four dimensions there are two
infilling Schwarzschild-AdS solutions in a finite isothermal cavity
with positive and negative specific heats. It remains to see how this
picture changes in higher dimensions and whether explicit solutions
for infilling geometries are possible. These will be reported
elsewhere.

We show that for all infilling solutions above, including those
occurring in pairs, the boundary is necessarily convex. This answers
the question posed by one of the authors at the Samos Meeting on
Cosmology, Geometry and Relativity, 1998 \cite{GCon} of whether
convexity can be applied as a criteria for selecting solutions if the
boundary admits more than one infilling geometry (of the same
type). We show that convexity cannot play such a role for Ricci-flat
spaces admitting $U(1)$ actions in general.  Using a relatively recent
theorem by Reilly \cite{Reilly}, we find that convexity gives a lower
bound of the Euclidean action for each infilling solution through the
$n$-volume of $\Sigma$ and the $(n+1)$-volume of the infilling
solutions ${\cal{M}}$.

Finally we have discussed the analogues of Minkowski's inequality and
found some interesting results. They can be grouped into two
classes. In one class we have the Taub-Nut and the Eguchi-Hanson
spaces. We found that the analogue of (flat-space's) Minkowski's
inequality for the Taub-Nut in arbitrary dimension is just the
opposite of Minkowski's inequality in that dimension. The
Eguchi-Hanson is found to lie in the middle in that its inequality has
the same {\it{sense}} as that of flat-space Minkowski and is
$\frac{1}{n+1}$ times the latter in $(2n+2)$ dimensions. We have
explained why this happens. The other class consists of the
Schwarzschild and the Taub-Bolt. In either case, the two infilling
geometries, although topologically equivalent, are strikingly
dissimilar in their isoperimetric inequalities. We have been able to
find the inequalities explicitly in the case of the Schwarzschild
solutions. The two inequalities are ``connected'' in that one is given
by being equal or greater than some value and the other by the
opposite of this inequality. Interestingly the isoperimetric
inequality for the larger mass stable black hole solution which has a
positive specific heat is always within Minkowski's bound. In
gravitational thermodynamics hot flat space is taken as the background
for the Schwarzschild calculations and hence this observation may have
some important thermodynamic consequences. For the Taub-Bolt the
nature of the polynomials involved prohibits one to obtain the
inequalities exactly, and also the two inequalities are not
``connected''. They both, however, have the same {\it{sense}} in that
they are given by being equal or less than some values. We have
obtained analytic expressions in a form from which it is
straightforward to find the approximate inequalities (for any choice
of base manifold $M$ and for any value of the periodicity $k$)
numerically should one need to know them. Note that both of the
inequalities obey the bound provided by their nut counterpart, i.e.,
by the Taub-Nut. Therefore the above comments about the flat space and
the Schwarzschild equally apply for the Taub-Nut and Taub-Bolt
solutions.

\subsection*{Acknowledgements}
We would like to thank John Barrow, Don Page, Simon Ross, Marika
Taylor and James York for useful remarks. Special thanks belong to
Christoph Galfard and Sakura Schafer-Nameki for translation. MMA was
supported by awards from the Cambridge Commonwealth Trust, the
Overseas Research Scheme and DAMTP.

\vskip1cm
\appendix
\section{Trinomial Equations}
In this appendix we describe the solution to the general trinomial
equation following Birkeland \cite{Birkeland}. We write the generic
trinomial equation in the form
\begin{equation}
x^{n}-ax^{s}+b=0 
\end{equation}
where $a$ and $b$ are coefficients, in general complex, and
$n>s$. Define
\begin{equation}
\zeta=\frac{n^{n}}{s^{s}(n-s)^{n-s}}\frac{b^{n-s}}{a^{n}}.
\end{equation}
Two possibilities can occur depending on whether $|\zeta|<1$ or
$|\zeta|>1$.
\section*{$|\zeta|<1$}
Define
\begin{equation}
x=z a^{\frac{1}{n-s}},\,\,l=-b a^{\frac{-n}{n-s}}
\end{equation}
The $n-s$ roots are given by
\begin{equation}
x^{\gamma}_{i}=a^{\frac{\gamma}{n-s}}
\epsilon^{i\gamma}\left(F_{0}(\zeta)+\frac{\gamma}{n-s}
\sum_{\kappa=1}^{n-s-1} \epsilon^{-in\kappa}\mu_{\kappa} l^{\kappa}
F_{\kappa}(\zeta)\right) \label{ns}
\end{equation}
\begin{displaymath}
(i=1,2,...,n-s)
\end{displaymath}
where $\gamma$ is an arbitrary constant and
\begin{equation}
\epsilon=e^{{\frac{2\pi}{n-s}}\sqrt{-1}},\,\,
\mu_{1}=1,\,\,\,\mu_{\kappa}=\frac{1}{\kappa}\left(\begin{array}{c}
\frac{\gamma -s \kappa}{n-s}-1\\ \kappa -1
\end{array}\right)
\end{equation}
and
\begin{equation}
F_{\kappa}(\zeta)=F \left(
\begin{array}{cccc}
a_{1,\kappa}, & \dots, & a_{n-1,\kappa}, & a_{n,\kappa}\\
b_{1,\kappa}, & \dots, & b_{n-1,\kappa}, & \zeta
\end{array}
\right)
\end{equation}
are the higher order hypergeometric functions of order $n$ in
general\footnote{When $\zeta=1$, $F_{\kappa}(\zeta)$'s reduce to
hypergeometric functions of order $n-1$.}:
\begin{equation}
F_{\kappa}(\zeta)=\sum^{\infty}_\mu \frac{(a_{1,\kappa})_{\mu}\dots
(a_{n,\kappa})_{\mu} }{(b_{1,\kappa})_{\mu}\dots (b_{n,\kappa})_{\mu}}
\frac{\zeta^{\mu}}{\mu!}
\end{equation}
in which the Pochhammer symbol has been used: $(a)_{\mu}$ stands for
$a(a+1)(a+2)\dots(a+\mu-1)$. Also here
\begin{equation}
\begin{array}{rcl}
a_{i,\kappa}&=&\frac{\kappa}{n-s}+\frac{n-i}{n}-\frac{\gamma}{n(n-s)}
\,\,\, (i=1,2,\dots,n),\\
b_{i,\kappa}&=&\frac{\kappa}{n-s}+\frac{s-i+1}{s}-\frac{\gamma}{s(n-s)}
\,\,\, (i=1,2,\dots,s),\\
b_{i,\kappa}&=&\frac{\kappa}{n-s}+\frac{i-s}{n-s}+\frac{\delta_{i}}{n-s}
\,\,\, (i=s+1,\dots,n-1).
\end{array}
\end{equation}
and 
\begin{equation}
\delta_{i}=0, {\mathrm{when}}\,\,\, i<n-\kappa,\,\,\delta_{i}=1,
{\mathrm{when}}\,\,\, i \ge n-\kappa.
\end{equation}
Now define 
\begin{equation}
x=z
(-l)^{\frac{\gamma}{s}}a^{\frac{1}{n-s}},\,\,\,\lambda=(-l)^{\frac{n-s}{s}}=b^{\frac{n-s}{s}}\,g^{-\frac{n}{s}}.
\end{equation}
Under the condition $|\zeta| <1$, the remaining $s$ roots are
\begin{equation}
x^{\gamma}_{n-s+i}=(-l)^{\frac{\gamma}{s}}g^{\frac{\gamma}{n-s}}
\delta^{i\gamma}\left[\phi_{0}(\zeta)+\frac{\gamma}{s}\sum_{\kappa=1}^{s-1}\delta^{i\,n\,\kappa}\,\Delta_{\kappa}\lambda^{\kappa}\phi_{\kappa}(\zeta)\right]
\label{s}
\end{equation}
\begin{displaymath}
(i=1,2,...,s),
\end{displaymath}
where
\begin{equation}
\delta =e^{{\frac{2\pi}{s}}\sqrt{-1}},\,\,\,\,\,\,
\Delta_{1}=1,\,\,\,\,\,\,\Delta_{\kappa}=\frac{1}{\kappa}\left(\begin{array}{c}
\frac{\gamma +n \kappa}{s}-1\\
\kappa -1
\end{array}\right)
\end{equation}
and
\begin{equation}
\phi_{\kappa}(\zeta)=F\left(
\begin{array}{cccc}
d_{1,\kappa}, & \dots, & d_{n-1,\kappa}, & d_{n,\kappa}\\
e_{1,\kappa}, & \dots, & e_{n-1,\kappa}, & \zeta
\end{array}
\right)
\end{equation}
in which
\begin{equation}
\begin{array}{rcl}
d_{i,\kappa}&=&\frac{\kappa}{s}+\frac{i-1}{s}+\frac{\gamma}{s\,n} \,\,\,
(i=1,2,\dots,n),\\
e_{i,\kappa}&=&\frac{\kappa}{s}+\frac{i}{n-s}+\frac{\gamma}{s\,(n-s)} \,\,\,
(i=1,2,\dots,n-s),\\
e_{i,\kappa}&=&\frac{\kappa}{s}+1+\frac{i-n}{s}+\frac{\delta_{i}}{s} \,\,\,
(i=n-s+1,\dots,n-1),
\end{array}
\end{equation}
and $\delta_{i}$ is the same as before.

The expansions (\ref{ns}) and (\ref{s}) converge for $|\zeta
|<1$. They will diverge for $|\zeta|>1$ and hence the letter is
treated separately.
\section*{$|\zeta|>1$}
The roots
for $|\zeta|>1$ are given by higher order hypergeometric function of
the variable $\frac{1}{\zeta}$. Defining
\begin{equation}
x=z (-b)^{\frac{1}{n}},\,\,\,l_{1}=\rho=g (-b)^{-\frac{n-s}{n}},
\end{equation}
all of the $n$ roots are given by
\begin{equation}
x^{\gamma}_{i}=\beta^{\frac{\gamma}{n}} \nu^{i\,\gamma}\left[
\psi_{0}\left( \frac{1}{\zeta}\right)+ \frac{\gamma}{n}
\sum_{\kappa=1}^{n-1}
\nu^{i\,s\,\kappa}\,\theta_{\kappa}\,\rho^{\kappa}\psi_{\kappa}\left(
\frac{1}{\zeta}\right)\right]\,\,\,\,\,\,\,\,\,(i=1,2,\dots,n),
\end{equation}
in which
\begin{equation}
\nu=e^{\frac{2\,\pi}{n}\sqrt{-1}},\,\,\,\,\,\,\,\,\,\theta_{1}=1,\,\,\,\,\,\,\,\,\,\theta_{\kappa}=\left(\begin{array}{c}
\frac{\gamma +s \kappa}{n}-1\\ \kappa -1
\end{array}\right),
\end{equation}
and $\gamma$ is an arbitrary constant as before.

The $\psi$'s here are hypergeometric functions of order $n$ with the
explicit form:
\begin{equation}
\psi_{\kappa}\left( \frac{1}{\zeta}\right)=F\left(
\begin{array}{cccc}
g_{1,\kappa}, & \dots, & g_{n-1,\kappa}, & d_{n,\kappa}\\
h_{1,\kappa}, & \dots, & h_{n-1,\kappa}, & \zeta
\end{array}
\right)
\end{equation}
in which
\begin{equation}
\begin{array}{rcl}
g_{i,\kappa}&=&\frac{\kappa}{n}+\frac{i-1}{s}+\frac{\gamma}{s\,n}
\,\,\, (i=1,2,\dots,s),\\
g_{i,\kappa}&=&\frac{\kappa}{n}+\frac{n-i}{n-s}-\frac{\gamma}{n\,(n-s)}
\,\,\, (i=1,2,\dots,n),\\
h_{i,\kappa}&=&\frac{\kappa+i}{s}+\frac{\delta_{i}}{n} \,\,\, (i=1,
2,\dots,n-1),
\end{array}
\end{equation}
where $\delta_{i}$ is the same as before.

\end{document}